\documentclass[12pt,a4paper]{article}
\usepackage{jheppub}
\pdfoutput=1
\usepackage{graphicx}
\usepackage{feynmp}
\newcommand{\fmfvcenter}[1]{\vcenter{\hbox{\fmfreuse{#1}}}}

\def\sgn{\text{sgn}}
\newcommand{\tl}[1]{\tilde{#1}}
\newcommand{\detr}{\text{det}\,}
\newcommand{\mc}[1]{\mathcal{#1}}
\def\[{\left[}
\def\]{\right]}
\def\({\left(}
\def\){\right)}
\def\cC{{\cal C}}

\def\cS{{\cal S}}

\newcommand{\p}{\partial}

\newcommand{\be}{\beta}

\def\sgn {\text{sgn}}

\def\Tr{\mathrm{Tr}}

\def\sgn{\text{sgn}}

\def\Tr{\mathrm{Tr}}

\def \be {\begin{equation}}
\def \ee {\end{equation}}
\def \bea {\begin{eqnarray}}
\def \eea {\end{eqnarray}}
\def \beal#1 {\begin{align}#1\end{align}}

\everymath{\displaystyle}
\newcommand{\td}{\mathcal{D}}
\newcommand{\mbb}[1]{\mathbb{#1}}

\preprint{TIFR/TH/18-10}

\title{Bose-Fermi Chern-Simons Dualities in the Higgsed Phase}

\author[a,1]{Sayantan Choudhury,\note{sayantan@aei.mpg.de,
    sayantan.choudhury@aei.mpg.de}} \author[b,2]{Anshuman
  Dey,\note{anshuman@theory.tifr.res.in}} \author[b,3]{Indranil
  Halder,\note{indranil.halder@tifr.res.in} } \author[c,4]{Sachin
  Jain,\note{sachin.jain@iiserpune.ac.in}} \author[b,5]{Lavneet
  Janagal,\note{lavneet@theory.tifr.res.in}} \author[b,6]{Shiraz
  Minwalla,\note{minwalla@theory.tifr.res.in}}\author[b,7]{Naveen
  Prabhakar \note{naveensp@theory.tifr.res.in}}
\affiliation[a]{Theoretical Cosmology Group, Max Planck Institute
  for Gravitational Physics, Albert Einstein Institute, Am
  M\"uhlenberg 1, 14476 Potsdam-Golm, Germany.}
\affiliation[b]{Department of Theoretical Physics, Tata Institute of Fundamental Research, Homi Bhabha Rd, Mumbai 400005, India}
\affiliation[c]{Indian Institute of Science Education and Research, Homi Bhabha Rd, Pashan, Pune 411 008, India}

\abstract{It has been conjectured that fermions minimally coupled to a
  Chern-Simons gauge field define a conformal field theory (CFT) that
  is level-rank dual to Chern-Simons gauged Wilson-Fisher Bosons. The
  CFTs in question admit relevant deformations parametrized by a real
  mass. When the mass deformation is positive, the duality of the two
  deformed theories has previously been checked in detail in the large
  $N$ limit by comparing explicit all orders results on both sides of
  the duality. In this paper we perform a similar check for the case
  of negative mass deformations. In this case the bosonic field
  condenses triggering the Higgs mechanism. The effective excitations
  in this phase are massive $W$ bosons. By summing all leading large
  $N$ graphs involving these $W$ bosons we find an all orders (in the
  't Hooft coupling) result for the thermal free energy of the bosonic
  theory in the condensed phase. Our final answer perfectly matches
  the previously obtained fermionic free energy under the conjectured
  duality map. }

\begin{document}

\maketitle

\section{Introduction}
It has recently been conjectured that fermions coupled to Chern-Simons
gauge theories in certain representations of the gauge group are dual
to bosons coupled, roughly speaking, to level-rank dual
representations of the level-rank dual Chern-Simons gauge theories.
The initial reason to suspect such a duality arose
\cite{Giombi:2011kc} from the study of conjectured dual bulk Vasiliev
duals of these theories \cite{Klebanov:2002ja, Sezgin:2002rt,
  Giombi:2009wh, Giombi:2011kc, Chang:2012kt}. Moreover in the papers
\cite{Giombi:2011kc, Maldacena:2011jn,Maldacena:2012sf} it was
demonstrated that Chern-Simons theories with matter in the fundamental
representation are `solvable' in the large $N$ limit. Using the
results of \cite{Maldacena:2011jn,Maldacena:2012sf} and the
Schwinger-Dyson techniques developed in \cite{Giombi:2011kc}), the
authors of \cite{Aharony:2012nh,GurAri:2012is} demonstrated that the
three point functions of single trace operators on the two sides of
the duality match at their conformal points provided the levels and
ranks of these theories are exchanged under the duality\footnote{See
  \cite{Bedhotiya:2015uga,Turiaci:2018nua} for further results on
  correlation functions.} and so provided the first concrete
conjecture for the including a map between dual parameters
\cite{Aharony:2012nh}\footnote{See
  \cite{Aharony:2015mjs,Seiberg:2016gmd,Karch:2016sxi,Murugan:2016zal}
  for more precise versions of the duality map.}.

While the matching of correlators between the two conformal theories
suggests that they are dual, this evidence alone is less than
clinching as the structure of large $N$ three point functions of
single trace operators in these theories is highly constrained by
approximate higher spin symmetries
\cite{Maldacena:2011jn,Maldacena:2012sf}. Compelling additional
evidence for these dualities - at least at large $N$ - comes from
matching thermal partition functions
\cite{Giombi:2011kc,Jain:2012qi,Yokoyama:2012fa, Aharony:2012ns,
  Jain:2013py,Takimi:2013zca,Yokoyama:2013pxa,Jain:2013gza,Minwalla:2015sca,xyz,Geracie:2015drf}
and S-matrices
\cite{Jain:2014nza,Dandekar:2014era,Inbasekar:2015tsa,Yokoyama:2016sbx,Inbasekar:2017ieo,Inbasekar:2017sqp}. It
turns out that these quantities can both be explicitly computed at
large $N$ independently in the bosonic and fermionic theories using
the techniques first introduced in \cite{Giombi:2011kc} and match
perfectly between bosons and fermions under the same duality map
proposed in \cite{Aharony:2012nh}, establishing the duality between
these theories beyond reasonable doubt, at least at leading order in
the large $N$ limit. The authors of \cite{Jain:2013gza, xyz} were also
able to construct pairs of dual \cite{Benini:2011mf} RG flows that
originate at the $\mc{N}=2$ supersymmetric field theory that terminate
in the IR at the critical boson and regular fermion theories
respectively. The fact that the supersymmetric duality of
\cite{Benini:2011mf} holds at finite $N$ supplies evidence for the
validity of Bose-Fermi dualities at finite (if large) $N$\footnote{See
  \cite{Aharony:2011jz,GurAri:2012is,Chang:2012kt,xyz,Jain:2012qi,Yokoyama:2012fa,Takimi:2013zca,Bardeen:2014paa,Bardeen:2014qua,Gurucharan:2014cva,Frishman:2014cma,Moshe:2014bja,Bedhotiya:2015uga,Geracie:2015drf,Minwalla:2015sca,Gur-Ari:2016xff,Giombi:2016ejx,Wadia:2016zpd,Giombi:2016zwa,Nosaka:2017ohr,Giombi:2017txg,Charan:2017jyc,abcd}
  for other large-$N$ computations that provide additional evidence for
  this duality.}.

Chern-Simons coupled regular fermion and critical boson theory CFTs
each admit a massive deformation labelled by a real mass, that map to
each other under duality. Turning on this mass deformation triggers an
RG flow. Positive and negative mass deformations both lead to gapped
theories or more precisely pure Chern-Simons topological field
theories (TFT)s. The low energy TFTs are different for the two signs
of mass. A change in sign of the fermion mass from positive to
negative decreases the level of the effective low energy topological
Chern-Simons theory by one unit. On the other hand, changing the boson
mass changes from positive to negative is expected to cause the boson
to condense and so to reduce the rank of the low energy topological
Chern-Simons theory by one unit. These two effects map to each other
under level-rank duality \cite{Aharony:2012nh}. In the large $N$
limit there is already considerable direct calculational evidence for
the duality between the two CFTs after a mass deformation for one sign
of the mass, as we now briefly review\footnote{Of course a duality
  between two CFTs implies a duality between dual pairs of relevant
  deformations of these CFTs. So the matching of physical quantities
  after deformation can be regarded as strong additional evidence for
  the duality between the parent CFTs.}.

Correlation functions are much more constrained at fixed points than
along RG flows. As a consequence there have been no exact results for
correlation functions in the mass deformed bosonic and fermionic
theories. It turns out, however, that the thermal partition functions
and S-matrices of both theories are roughly as easy to compute at
large $N$ in the mass deformed theories (for both signs of the
fermionic mass and for positive bosonic mass) as at the fixed point.
Explicit all-orders results for the partition function and S-matrix
are already available for positive bosonic masses and fermionic masses
of both signs; and - to the extent that they can be compared - match
perfectly under duality.
\cite{Giombi:2011kc,Jain:2012qi,Yokoyama:2012fa, Aharony:2012ns,
  Jain:2013py,Takimi:2013zca,Yokoyama:2013pxa,Jain:2013gza,Minwalla:2015sca,xyz,Geracie:2015drf}. However,
explicit all-orders results for the negative mass deformed bosonic
theory have not been as easy to obtain. At the calculational level a
negative bosonic mass causes the bosons to condense, completely
changing the nature of the mathematical problem to be solved. In order
to determine the free energy and S-matrices of the bosonic theory in
the condensed phase, in other words, one is required to solve a new
mathematical problem that is not a small deformation of the analogous
problem solved at the conformal fixed point.

In this paper we solve this `new mathematical problem' for the bosonic
free energy at finite temperature, and thereby present an all-orders
`solution' of the bosonic theory with a negative mass deformation. We
proceed as follows. First we note that the effective excitations in
the Higgsed bosonic phase are $W$ bosons. We work in a mixed unitary
- lightcone gauge that is convenient for our problem, and reduce the
original scalar Chern-Simons Lagrangian to a Lagrangian that describes
the interaction of the charged $W$ bosons with the unbroken part of
the gauge group. We then use Schwinger-Dyson methods to sum all
diagrams that contribute at leading order in the large $N$ limit to
the thermal propagator of the W bosons. The free energy is then
obtained by sewing this exact thermal propagator on itself and adding
in some appropriate counterterms. Our final result for the thermal
free energy (and thermal mass of the $W$ bosons) turns out to match
perfectly with the previously obtained dual fermionic results.
 
We leave the generalization of the computations of this paper to
S-matrices to future work. We also work only with critical bosons,
leaving the generalization to other theories to future work. We also
work only in the strict large $N$ limit. See
\cite{Aharony:2015pla,Radicevic:2015yla,Aharony:2015mjs,Hsin:2016blu,
  Radicevic:2016wqn,Karch:2016aux,Seiberg:2016gmd,Aharony:2016jvv,Karch:2016sxi,Benini:2017dus,Gaiotto:2017tne,Jensen:2017dso,Jensen:2017xbs,Gomis:2017ixy,Cordova:2017vab,Benini:2017aed,Jensen:2017bjo}
for exciting recent progress on the study of these dualities, 
and additional checks of the dualities, at finite
$N$.

In the rest of this introduction we provide a more detailed
description of the theories we study, our explicit results in the
context of what was already known, as well as the interesting physical
implications of our computations.

\subsection{Theories and conjectured dualities}

The two classes of theories we study in this paper are Regular Fermion (RF) theories defined by the Lagrangian 
\begin{equation}
 S_{\text{RF}}[\psi] = S_{\text{CS}} + \int d^3 x  \left( \bar{\psi} \gamma_\mu D^{\mu} \psi  + m_F^{\text{reg}} \bar \psi \psi \right)\ ,
\label{rft1}
\end{equation}
and the so called Critical Boson (CB) theory defined by the Lagrangian 
\begin{equation}
S_{\text{CB}}[\phi, \sigma_B]  = S_{\text{CS}}
 + \int d^3x \left[ D_\mu \bar \phi D^\mu\phi+ \sigma_B \left(\bar\phi \phi  +\frac{N_B}{4\pi} m_B^{\text{cri}} \right) \right]\ .
\label{cst1}
\end{equation}

In the actions \eqref{rft1} and \eqref{cst1}, $S_{\text{CS}}$ denotes
the pure gauge action for three dimensional gauge fields which is of
the pure Chern-Simons form without admixture of a Yang-Mills
term\footnote{One way of giving precise meaning to the theories
  studied in this paper is by turning on a Yang Mill term with a small
  gauge coupling and then taking the limit in which this coupling goes
  to zero.}. The field $\sigma_B$ in \eqref{cst1} is a Lagrange
multiplier field\footnote{One way of thinking of this field is as a
  Hubbard-Stratonovich field that accounts for a $\phi^4$ interaction
  between the bosons. In the limit that the $\phi^4$ coupling becomes
  very large this quadratic term for the Hubbard-Stratonovich field
  vanishes, turning it into a Lagrange multiplier.} and its presence
is the manifestation of the fact that we are studying the `critical'
or gauged Wilson-Fisher scalar theory rather than the regular or
gauged free scalar theory.

In this paper we restrict our attention to the gauge groups $SU(N_F)$
or $U(N_F)$ (for the fermionic theory) and $SU(N_B)$ or $U(N_B)$ for
the bosonic theory. The Chern-Simons level for the $SU(N_B)$
Chern-Simons action in the bosonic theory will be denoted by the
integer $k_B$\footnote{In the case that the bosonic theory is
  $U(N_B)$, the rank of its $U(1)$ part is either $N_B k_B$ - in the
  case of the so called type II theory or
  $N_B {\sgn( k_B)} (|k_B|+ N_B)$ in the case of the so called type 1
  theory. See \cite{Radicevic:2016wqn} for a generalization 
  of the type 1 and type II theories to a more general set of 
  so called $(k, k')$ theories. }. We define levels of the fermionic theory to be the level
of the pure Chern-Simons theory obtained in the IR by giving the
fermions a mass of the same sign as their level and integrating them
out. With this definition, the rank of the $SU(N_F)$ part of the
fermionic theory is denoted by $k_F$\footnote{Once again, the case
  that the fermionic theory is $U(N_F)$, the rank of its $U(1)$ part
  is either $N_F k_F$ - in the case of the so called type II theory or
  $N_B {\sgn( k_F)} (|k_F|+ N_F)$ in the case of the so called type 1
  theory.}. In the large $N$ limit the $SU(N)$ theories and the two
$U(N)$ theories all coincide, so we effectively treat them as
identical and deal with them all together in the computations
presented in this paper.

It has been conjectured that the fermionic and bosonic theories
described above are dual to each other when their levels and ranks are
related as follows
\begin{equation}\label{lod}
k_F= -{\rm sgn}(k_B) N_B\ , ~~~ N_F= |k_B|\ . 
\end{equation}
In the large $N$ limit of these theories, instead of working with
levels and ranks, it is useful to parametrize these theories by
their `renormalized' levels $\kappa_{F}$, $\kappa_B$ and 't Hooft
couplings $\lambda_{F}$, $\lambda_B$ defined by
\begin{equation}\label{rl}
\kappa_B= {\rm sgn}(k_B)( |k_B| +N_B), ~~~\kappa_F={\rm sgn}(k_F) (|k_F|+N_F), ~~~\lambda_F= \frac{N_F}{\kappa_F}, ~~~\lambda_B= \frac{N_B}{\kappa_B}\ .
\end{equation}
In terms of these variables, the conjectured duality map may be stated
as
\begin{equation}\label{dm}
\kappa_F= -\kappa_B\ , ~~~\lambda_F=- {\rm sgn}(\lambda_B) +\lambda_B\ .
\end{equation}
At least in the large $N$ limit, the conjectured duality map for mass
parameters between the bosonic and fermionic theories is given by
\begin{equation}\label{cmmp}
m_F^{\text{reg}}= -\lambda_B m_B^{\text{cri}}\ .
\end{equation}

\subsection{Recap of known results}
\subsubsection{Structure of the large \texorpdfstring{$N$}{N} partition function} \label{holint}
It was demonstrated in \cite{Jain:2013py} that the partition function
of theories \eqref{rft1} and \eqref{cst1} on $S^2 \times S^1$ can be
evaluated (in a suitably coordinated high temperature and large $N$
limit) by following a two step procedure that we now outline.

In the first step we are instructed study the theory in question on
$\mbb{R}^2 \times S^1$. Up to gauge transformations, the zero mode of
the holonomy $U$ of the gauge field around $S^1$ is completely
specified by its eigenvalues $e^{i \theta_j}$ where $j= 1 \ldots N$
and
$$\theta_j \in  (-\pi, \pi]\ . $$ 
In the large $N$ limit, all information about the eigenvalues of the
holonomy matrix is conveniently packaged into an eigenvalue
distribution function $\rho(\alpha)$ defined by
\begin{equation}\label{rhotet}
\rho(\alpha)= \frac{1}{N} \sum_{i=1}^N \delta(\alpha-\theta_i) 
\end{equation}
To complete the first step we are instructed to evaluate the path
integral of our theory at fixed values of the holonomy zero mode
$U$. This path integral defines the `free energy functional'
$v[U]$\footnote{We alternately use the notation $v[\rho]$ to denote
  the dependence of $v$ on the holonomy eigenvalue distribution
  $\rho(\alpha)$ \eqref{rhotet} as is appropriate in the large $N$
  limit.  } via the schematic equation
\begin{equation}\label{seff}
  e^{-\mc{V}_2 T^2 v[U] } = \int_{\mbb{R}^2 \times S^1} [d \phi]\ e^{-S[\phi, U]}\ .
\end{equation}
where $\mc{V}_2$ is the volume of two dimensional space and $T$ is the
temperature. In order to complete the evaluation of the partition
function of interest, we are instructed to evaluate the unitary matrix
integral
\begin{equation}\label{ioev}
\mc{Z}_{S^2\times S^1} =\int [dU]_{\text{CS}} ~e^{-\mc{V}_2 T^2 v[U]}.
\end{equation}
(this is the second step of our procedure). Here $[dU]_{\text{CS}}$
is a Chern-Simons modified Haar measure over $U(N)$ (see
\cite{Jain:2013py} for full details). In the large $N$ limit under
study, \eqref{ioev} is most conveniently evaluated in the saddle point
approximation.

In this paper we focus entirely on the first step of this programme,
i.e. the evaluation of $v[U]$ (or more appropriately $v[\rho]$)
defined in \eqref{seff}. In order to investigate the actual physics of
our theories (as opposed to simply verifying duality) one would have
to study the matrix integral \eqref{ioev}; we leave this to future work.

\subsubsection{Free energies and gap equations}
We now review what is already known about the functionals $v_F$ and
$v_B$ i.e.~$v$ evaluated using \eqref{seff} starting with the
fermionic theory \eqref{rft1} and the bosonic theory \eqref{cst1}
respectively.

By explicitly evaluating an infinite sum of loop diagrams of the
fermionic/bosonic fields one obtains off-shell free energy functionals
$v_F(|c_F|, \rho_F)$ and $v_B(|c_B|, \rho_B)$ for these two theories
respectively. Note that each of these off shell free energies depend
on an auxiliary variable (i.e. $|c_F|$ or $|c_B|$) in addition to the
holonomy eigenvalue distribution functions. The onshell free energy
functionals defined in \eqref{seff} are functions only of the holonomy
fields; and are obtained from their off-shell counterparts by
extremizing w.r.t.~$|c_F|$ and $|c_B|$ respectively. The extremum
values of $|c_F|$ and $|c_B|$ are physically interpreted as thermal
pole masses in units of the temperature for the fermionic/bosonic
theories respectively. In case the off-shell free energy admits more
than one extremum, we are instructed to choose the extremum with the
lowest value of the free energy. The explicit results for the
fermionic and bosonic off-shell free energies are
\begin{equation} \label{offshellfe}
\begin{split}
v_B(|c_B|, \rho_B) &=\frac{N_B}{6\pi} {\Bigg[}\frac{3}{2} {\hat m}_B^{\text{cri}} c_B^2- \frac{1}{2}\left({\hat m}_B^{\text{cri}}\right)^3  -|c_B|^3 + \\
&  +3 \int_{-\pi}^{\pi} \rho_B(\alpha) d\alpha\int_{|c_B|}^{\infty}dy y\left(\log\left(1-e^{-y-i\alpha -\nu}\right)+\log\left(1-e^{-y+i\alpha + \nu}\right)  \right)
 {\Bigg]},\\
 v_F(|c_F|,\rho_F) &=\frac{N_F}{6\pi} {\Bigg[}  |c_F|^3 \frac{\left(\lambda_F-\sgn(X_F)\right)}{\lambda_F} +\frac{3}{2 \lambda_F} {\hat m}_F^{\text{reg}} c_F^2- \frac{\left({\hat m}_F^{\text{reg}}\right)^3}{2\lambda_F \left(\lambda_F-\sgn(X_F)\right)^2}\\
&-3 \int_{-\pi}^{\pi} \rho_F(\alpha) d\alpha\int_{|c_F|}^{\infty}dy y\left(\log\left(1+e^{-y-i\alpha -\nu}\right)+\log\left(1+e^{-y+i\alpha + \nu}\right)  \right)
 {\Bigg]}\ .
\end{split}
\end{equation}
Here, ${\hat m}_B^{\text{cri}}$ and ${\hat m}_F^{\text{reg}}$ are the
masses divided by the temperature, and $\nu=\mu/T$ where $\mu$ is a
chemical potential that couples to a charge under which all
fundamental fields have unit positive charge while all antifundamental
fields have unit negative charge. The quantity $X_F$ 
(and an analogous quantity $X_B$ that we will need in a moment) that appears in \eqref{offshellfe} are defined by 
\begin{equation}\label{XFXC}
\begin{split}
 X_F&=2\lambda_F \cC+{\hat m}_F^{\text{reg}}\ , \\
 X_B&= 2 \lambda_B \cS -\lambda_B \hat{m}_B^{\text{cri}} -{\rm sgn} (\lambda_B){\rm max}(|c_B|, |\nu|)\ ,
\end{split} 
\end{equation}
where
\begin{equation}\begin{split}
\cC(|c_F|, \nu) =&\frac{1}{2} \int d\alpha\rho_F(\alpha)   \( \log(2 \cosh (\tfrac{|c_F| +i\alpha + \nu }{2}))+ \log(2 \cosh (\tfrac{|c_F| - i\alpha - \nu}{2})) \),\\
\cS(|c_B|, \nu) =&\frac{1}{2}\int d\alpha\rho_B(\alpha)   \( \log(2 \sinh (\tfrac{|c_B| +i\alpha + \nu}{2}))+ \log(2 \sinh (\tfrac{|c_B| - i\alpha -\nu}{2})) \).
\label{ss}
\end{split}\end{equation}

The expression for $v_F$ listed in \eqref{offshellfe} is expected to
be complete. However the expression for $v_B$ was obtained by working
about an unHiggsed bosonic vacuum. As we have explained earlier in
this introduction, under certain circumstances - roughly for negative
bosonic mass - we expect this to be the wrong vacuum for the bosonic
theory. As a consequence we expect the expression for $v_B$ above to
be correct only in a certain parametric regime. In this paper we will
calculate the bosonic free energy in the complementary parametric
regime\footnote{Our result for the free energy \eqref{offshellfe}
  is given in terms of the logarithmic function $\log (z)$. Here and
  through the rest of this paper this function is defined so that it
  is real when $z$ is real and positive, and so that it has a branch
  cut on the negative real axis in the $z$ plane.}.

The condition that $v_F$ and $v_B$ are extremized on-shell yields 
an equation -- called a \emph{gap equation} -- that can be used to
determine $c_F$ and $c_B$ on-shell. The gap 
equations for the fermionic theory is 
\begin{equation}\label{mdrfi}
|c_F|=\sgn(X_F) \left( 2 \lambda_F \cC(|c_F|, \nu)  +{\hat m}_F^{{\rm reg}}\right) = |X_F|\ ,
\end{equation}
while the gap equation obtained by extremizing 
$v_B$ is 
\begin{equation}\label{mdcbi}
2 \cS(|c_B|,\nu) = {\hat m}_B^{{\rm cri} }\ ,
\end{equation}
In Appendix \ref{review} below we review some properties of the gap equation. In particular we demonstrate that the bosonic
gap equation \eqref{mdcbi} only has solutions provided
\begin{equation}\label{dlcbinta}
  -\sgn(\lambda_B)\sgn(X_B) \geq 0\ .
\end{equation}
\footnote{\eqref{dlcbinta} should be thought of as effectively
  describing the parametric regime in which the bosonic free energy
  listed in \eqref{offshellfe} is valid. Outside this regime the free
  energy expression is obtained by the computation in the Higgsed
  phase that we perform in this paper.}When transformed to fermionic
variables, \eqref{dlcbinta} turns into the condition
\begin{equation}\label{dlcb}
\sgn(\lambda_F)\sgn(X_F) \geq 0\ .
\end{equation}

It turns out (see Appendix \ref{review}) that there is a one to one map
between all solutions to the bosonic gap equation \eqref{mdcbi} and and those solutions to the fermionic gap
equation \eqref{mdrfi} that obey \eqref{dlcb}. 

However there also exist solutions to
the fermionic gap equation that obey the complement of \eqref{dlcb}.The bosonic duals of these solutions have not been understood in the existing literature. However the duality map 
(see Appendix \ref{review} for details) can be used to recast
the existing fermionic results into bosonic variables, yielding 
the gap equation 
\begin{equation}\label{mdrfooi}
2|c_B| =\left( 2|\lambda_B| {\tilde \cS}
-|\lambda_B| \hat{m}_B^{\text{cri}}\right)\ .
\end{equation}
where 
\begin{equation}\label{deftcsi}
{\tilde \cS}= \left\{\begin{array}{cc} \cS  & {{\rm when} ~~|\nu|<|c_B|}\\ \cS - \frac{1}{2 |\lambda_B|} (|\nu|-|c_B|) & {{\rm when} ~~|\nu|>|c_B|}\end{array}\right.\ .
\end{equation}
In this range of parameters the off-shell fermionic free energy can also be recast into bosonic language; we find 
\begin{align} \label{prdfi}
&v_B(|c_B|,\rho_B)=\frac{N_B}{6\pi} {\Bigg[}-\frac{|\hat{m}|^3}{|\lambda_B| }+3|\lambda_B|   |\hat{m}| \cS^2+2|\lambda_B| ^2  \cS^3 \nonumber\\
&\qquad -|c_B|^3 + 3 \int_{-\pi}^{\pi} \rho_B(\alpha) d\alpha\int_{|c_B|}^{\infty}dy y\left(\log\left(1+e^{-y-i\alpha -\nu}\right)+\log\left(1+e^{-y+i\alpha + \nu}\right)\right)   {\Bigg]}\ ,
\end{align}

The equations \eqref{mdrfooi} and \eqref{prdfi} may be thought of as
the predictions of duality for the gap equation and free energy of the
bosonic theory in the condensed phase, i.e. when \eqref{dlcbinta} is
not obeyed. In this paper we will reproduce both these results from a
direct evaluation of the bosonic path integral in the condensed
phase. The exact agreement under duality of the bosonic and fermionic
results may be thought of as a detailed new check of the duality
conjecture.

\subsection{Computations in the Higgsed phase of the bosonic theory}

Consider the theory \eqref{rft1} when $m_B^{\text{cri}}<0$. In this
situation the equation of motion for the field $\sigma_B$ forces the
modulus of ${\bar \phi} \phi$ to take a fixed nonzero constant value
determined by $m_B^{\text{cri}}$. It is useful to work in the unitary
gauge. This choice of gauge rotates $\phi$ to be real and to point in
the $N_B$-th flavour direction in $SU(N_B)$ colour space\footnote{For
  definiteness sake, we work with the gauge group being $SU(N_B)$ as
  opposed to $U(N_B)$.}. As $|\phi|$ is a fixed constant,  it follows that with this choice of gauge there
are no degrees of freedom in $\phi$.

The `condensation' of $\phi$ breaks $SU(N_B)$ to
$SU(N_B-1)$. The gauge bosons of $SU(N_B)$ split up into Chern-Simons
coupled $SU(N_B-1)$ gauge bosons, $N_B-1$ complex massive $W$ bosons
that transform in the fundamental of $SU(N_B-1)$ and a single real
massive $Z$ boson that is uncharged under $SU(N_B-1)$. The action that
governs the interactions of these fields is easily worked out and is
presented in \eqref{lcgmom}, \eqref{lfmd} below.

As in previous work (see \cite{Giombi:2011kc} and several 
subsequent papers) we choose to adopt the lightcone gauge for the
unbroken $SU(N_B-1)$ gauge symmetry. Once we make this choice the
Lagrangian is quadratic both in the $SU(N_B-1)$ gauge fields and in
the $Z$ bosons. It is thus possible to integrate both these fields
out. This process generates a non-local quartic interaction between the
$W$ bosons. The resultant effective theory usually is called a vector
model in the literature on large $N$ models. The $W$ fields are
$SU(N_B-1)$ vectors with a non-local $W^4$ interaction. At this point
we use standard large $N$ techniques to reduce the finite temperature
path integral over $W$, in the large $N$ limit, to a set of nonlinear
integral equations for the self-energy matrix $\Sigma^{\mu\nu}$ of the
$W_\mu$ fields.

Using symmetries and other structural properties of the integral
equations, it is possible to parametrize the four nonzero components
of $\Sigma^{\mu\nu}$ by four unknown functions $F_1, \ldots, F_4$ of a
single variable (see \eqref{qitfoff}). The integral equations for
$\Sigma^{\mu\nu}$ reduce to a set of four nonlinear coupled integral
equations for $F_1, \ldots, F_4$ (see \eqref{inteqexp}). Quite
remarkably it turns out to be possible to solve these equations
exactly, in terms of a single real constant $M$. $M$ is the thermal
mass of the $W$ bosons and it, in turn, is required to obey a gap
equation \eqref{fge} which perfectly reproduces the prediction from
duality \eqref{mdrfooi}, at least when $|c_B|>|\nu|$.

Next we plug our solution for $\Sigma^{\mu\nu}$ back into our large
$N$ expression for the partition function. On-shell, we find that the
final result for the free energy $v_B$ perfectly matches the
prediction \eqref{prdfi}.

\section{Finite Temperature Partition function at large \texorpdfstring{$N_B$}{NB}}
\subsection{The effective action in terms of \texorpdfstring{$W$}{W} bosons}\label{Large N Effective Action}
Consider the mass deformed $SU(N_B)$ critical boson theory defined by the Euclidean action $S_{\text{E}}$
\footnote{We follow the standard convention that the Euclidean partition function is given 
in terms of the Euclidean action by the path integral $\int [d \phi]\, e^{-S_{\text{E}}[\phi]}$.}
\begin{equation} \label{basiclag}
\begin{split}
	S_{\text{E}} &=S_{\text{CS}} + S_{\text{B}}\ ,\\
	S_{\text{CS}} & =\int d^3x \ i \epsilon^{\mu \nu \rho} \frac{\kappa_B}{4 \pi}\, \Tr(X_\mu \partial_\nu X_\rho-\frac{2 i}{3} X_\mu X_\nu X_\rho)\ ,\\
	S_{\text{B}} &=\int d^3x \sqrt{\det g} \left(D_\mu\bar{\phi}D^\mu \phi+  \sigma_B \left( {\bar \phi} \phi + \frac{N_B}{4 \pi} m_B^{\text{cri}}\right)\right)\ .
\end{split}
\end{equation}
where $D_\mu \phi = \partial_\mu\phi -i X_\mu \phi$. The fields
$X_\mu$ are $N \times N$ hermitian matrices. Throughout this paper the
gauge group generators are normalised such that
$Tr(T_A T_B)=\tfrac{1}{2}\delta_{AB}$. All through this paper we work
with the dimensional regulation scheme (see subsection \ref{dimreg}
below). With this choice of regulator the constant $\kappa_B$ that
appears in \eqref{basiclag} matches $\kappa_B$ defined in
\eqref{rl}. In other words $\kappa_B$ may be identified with the
`renormalized' Chern-Simons level (see the discussion above
\eqref{rl}). We will present most of our formulae in terms of the 't
Hooft coupling defined by $\lambda_B=\frac{N_B}{\kappa_B}$ (see
\eqref{rl}).  Note that, by definition, $|\lambda_B|\leq 1$.

The field $\sigma_B$ plays the role of a Lagrange multiplier in the
action \eqref{basiclag}. The $\sigma_B$ equation of motion
\begin{equation}\label{phbb}
 {\bar \phi}{\phi}= - \frac{N_B}{4 \pi} m_B^{\text{cri}} \ ,
\end{equation}
has no real solution when $m_B^{\text{cri}}$ is positive. In this
`standard' case one proceeds to analyse the theory assuming that the
scalar field has a vacuum at $\phi=0$ and tests the self consistency
of this assumption by demonstrating that the quantum effective action
for $\sigma_B$ - evaluated by integrating out the $\phi$ fields - has a
stable minimum.

In this paper we are interested in regime in which $m_B^{\text{cri}}$
is negative. In this case the situation is more straightforward. Let
us define the real quantity $v$ by the equation
\begin{equation}\label{defv}
|\kappa_B| v^2=- \frac{N_B}{4 \pi} m_B^{\text{cri}} \implies v^2=- \frac{|\lambda_B|}{4 \pi} m_B^{\text{cri}}
\end{equation}
a definition that is sensible precisely because $m_B^{\text{cri}}$ is negative. 
The equation \eqref{phbb} may be rewritten as
\begin{equation}
 {\bar \phi}{\phi}= |\kappa_B| v^2\ .
\end{equation}
Clearly \eqref{phbb} now admits classical solutions, but the solution
to this equation is not unique; given any solution one can always
generate a new solution by performing a spacetime dependent $SU(N_B)$
rotation of the original solution. Of course these rotations are
simply gauge transformations, the solution to \eqref{phbb} can be made
unique by making an appropriate choice of gauge.  We adopt the so
called `unitary gauge' in which the fundamental field $\phi$ is always
rotated to be real and to point in the $N_B$-th direction in colour
space.  The equation of motion \eqref{phbb} then determines the
magnitude of $\phi$ at each point in spacetime and we find
\begin{equation} \label{sqv}
\phi^i  = \delta^{i N_B} v \sqrt{|\kappa_B|}= \delta^{i N_B} \sqrt{\frac{N_B}{4 \pi} |m_B^{\text{cri}}|} \ .
\end{equation}
With this choice the $\phi$ field is completely determined and
effectively non dynamical, and the original $SU(N_B)$ gauge symmetry
is Higgsed down to $SU(N_B-1)$.

Matter Chern-Simons theories are often thought of as theories which
govern interaction of dynamical matter fields with non-dynamical gauge
fields. In the current situation our gauge choice has frozen the
matter field completely. So where has its degrees of freedom gone? Of
course the answer to this question is familiar; symmetry breaking of
the gauge group transfers the degrees of freedom of the matter field
into the gauge field. More precisely, those gauge bosons that
originally were in the adjoint of $SU(N_B)$ - but are not in the
adjoint of the residual $SU(N_B-1)$, inherit the matter degrees of
freedom and turn into propagating massive $W$ bosons after symmetry
breaking. We now explain in detail how this works.

Let $(X_\mu)^i_j$ represent the $ij^{\text{th}}$ element
of the matrix valued field $X_\mu$ in \eqref{basiclag}.  The indices
$i$ and $j$ run over the range $i, j=1, \ldots, N_B$. It is useful to
separate out $i=N_B$ as special (same for $j$). Let $a, b$ denote
indices that run from $1, \ldots, N_B-1$. Then we define
\begin{equation} \label{xdecomp}
 (X_\mu)^a_{N_B}= \frac{W_\mu ^a}{\sqrt{\kappa_B}}, ~~~(X_\mu)^{N_B}_b= \frac{({\bar W}_\mu)_b}{ \sqrt{\kappa_B}}, ~~~ (X_\mu)^{N_B}_{N_B}= Z_\mu, ~~~(X_\mu)^a_b= (A_\mu)^a_b - \frac{Z_\mu}{N_B-1} 
 \delta^a_b  ~~~~
\end{equation}
where the traceless matrices $(A_\mu)^a_b$ are the gauge bosons of the
unbroken $SU(N_B-1)$ gauge group and $(W^a_\mu)^*= ({\bar W}_\mu)_a$,
i.e.~the fields $W_\mu$ and ${\bar W}_\mu$ are complex conjugates of
each other. Notice that $W_\mu$ transforms in the fundamental while
${\bar W}_\mu$ transforms in the antifundamental of the unbroken
gauge group $SU(N_B-1)$.

Working in unitary gauge and with the decomposition described in
\eqref{xdecomp} we obtain the Euclidean action
\begin{align}\label{asb}
	S_{\text{E}}[A,W,Z]&=\frac{i \kappa_B}{4 \pi}\int \Tr(AdA-\frac{2i}{3}AAA)+\frac{i }{4 \pi}\int [2 \bar{W}DW+\kappa_B ZdZ - 2iZ\bar{W}W] \nonumber\\ 
                  &\quad +{\sgn}(\kappa_B) v^2\int d^3x \sqrt{\det g}\,(\kappa_B Z_\mu Z^\mu + \bar{W}_\mu W^\mu)
\end{align}
where $D_\mu = \partial_\mu -i A_\mu$ and the exterior product $ABC$ means $d^3x\, \epsilon^{\mu\nu \rho} A_\mu B_\nu C_\rho$. 

At the linearized order the equation of motion for the field $A_\mu$
is simply $F_{\mu\nu}=0$, reflecting the fact that the field $A_\mu$ has 
no propagating degrees of freedom. On the other hand the linearized
equations of motion for the fields $W_\mu$ and $Z_\mu$ are
\begin{equation}\label{leomwz}
  \frac{ i  \epsilon^{\mu \nu \rho}}{4 \pi}  2\partial_\nu W_\rho + {\sgn}(\kappa_B)v^2 W^\mu=0, ~~~ 
  \frac{i  \epsilon^{\mu \nu \rho}}{4 \pi} 2\partial_\nu Z_\rho + {\sgn}(\kappa_B) v^2 2Z^\mu=0
\end{equation}
It follows immediately from the divergence of \eqref{leomwz} that
$ \partial_\mu W^\mu = \partial_\mu Z^\mu=0$.  The equations
\eqref{leomwz} are easily solved in momentum space. Let us define
\begin{equation}
    W^a_\mu(x)=\int \frac{d^3q}{(2 \pi)^3} \ e^{i x\cdot q} W^a_\mu(q) \ .
\end{equation}
Then, \eqref{leomwz} turns into
\begin{equation}\label{eve1}
\left( - \frac{\epsilon^{\mu  \nu \rho} q_\nu}{2 \pi }  +  {\sgn}(\kappa_B) v^2 g^{\mu \rho} \right) W^a_\rho(q)=0
\end{equation} 
The equation \eqref{eve1} has solutions only when the matrix on the
LHS of \eqref{eve1} has a zero eigenvalue. This, in turn, is the case
only when
\begin{equation} \label{baremass}
 -g^{\mu\nu} q_\mu q_\nu = (2 \pi v^2)^2 \equiv m_W^2 
\end{equation}
It follows that the $W$-boson $W_\mu$ is a propagating field with mass
$m_W = +2\pi v^2$. For every $q_\mu$ (at fixed $v^2$) for which
\eqref{baremass} is obeyed, the solution to \eqref{eve1} is uniquely
determined up to a single complex number (either $W_+ $ or $W_-$ for
example). It follows, in other words, that the fields $W^a_\mu$ have
the same number of degrees of freedom as a standard massive scalar
field of mass $m_W$. In a similar way the $Z$ boson $Z_\mu$ has as as
many solutions as a single real scalar field of mass $2 m_W$. The
total number of degrees of freedom of the $W, Z$ system is thus that
of $2(N_B-1)+1=2N_B-1$ real massive scalars. This is precisely the
number of degrees of freedom in the scalar field once its modulus has
been frozen by the $\sigma_B$ equation of motion. The condensation of
$\phi$ simply `transmutes' these degrees of freedom from a spin zero
scalar to a spin $\pm 1$ vector\footnote{If we take the momentum of
  the $W$ boson to be in the $3$ direction, it is easy to check that
  its polarization is in the $z= x+iy$ direction when $\kappa_B$ is
  positive but in the ${\bar z} =x-iy$ direction when $\kappa_B$ is
  negative (this follows from \eqref{eve1} using \eqref{baremass}).
  In other words the little group spin of our on-shell $W$ bosons
  equals ${\rm sgn}(\kappa_B) $.}.

\subsection{Reducing the evaluation of the partition function to
  saddle point equations}
In the rest of this section we will evaluate the finite temperature
partition function 
\begin{equation}\label{ftpf}
\mathcal{Z} = \int [dA dW dZ]\ e^{-S_{\text{E}}[A,W,Z]}\ ,
\end{equation}
where the action $S_{\text{E}}$ was defined in \eqref{asb} and the path integral is 
evaluated over the Euclidean manifold $\mbb{R}^2 \times S^1$. 

Adapting the explanation of \cite{Jain:2013py} to the current context,
it is possible to convince oneself that the path integral \eqref{ftpf}
may be evaluated (in the coordinated high temperature and large $N_B$
limit described in \cite{Jain:2013py}) by following a two step
process. The first step in this process is to evaluate the path
integral over the fields of \eqref{asb} on $\mbb{R}^2 \times S^1$ at
fixed values of the holonomy fields of the unbroken gauge group
$SU(N_B-1)$. The second step in this process is to regard the result
of the first path integral as an effective action for the holonomy
matrix, and then use this action to perform an integral over the
holonomy matrices. Both steps in this process can be practically
carried through in the large $N_B$ limit (the second step involves
solving a saddle point equation for the holonomies). In this section
we concentrate entirely on the first step, leaving the second step for
later work.

In order to proceed with our computation we need to fix the unbroken $SU(N_B -1)$ gauge invariance in 
the action \eqref{asb}. Following \cite{Giombi:2011kc} (and more or less every other subsequent successful
large $N$ computation in matter Chern-Simons theories) we work in the lightcone
gauge $A_-=0$\footnote{Our conventions are defined by 
\begin{equation}
x^{\pm}= \frac{x^1\pm i x^2}{\sqrt{2}}, \ p_{\mp}=\frac{p^1\pm i p^2}{\sqrt{2}},\  A_{\mp}=\frac{A^1\pm i A^2}{\sqrt{2}}\ .
\end{equation} 
In our conventions, the nonzero components of the metric in lightcone
coordinates are $g_{+-}=g_{-+}= g_{33}=1$, the Levi-Civita symbol is
given by $\ \epsilon^{+-3}= \epsilon_{-+3}=-i\ .$ and the Kronecker
Delta is given by $\delta^{\mu}_{\nu}=1 \ \text{if} \ \mu=\nu $ and
$0$ otherwise.}.  Once we adopt this gauge, the cubic term for $A$
vanishes in the Chern-Simons action. The action \eqref{asb} simplifies
to
\begin{equation}\label{lcgauge}
	\begin{split}
	S_{\text{E}}[A,W,Z] &=\frac{i}{4 \pi}\int d^3x\ \text{Tr}\left(\kappa_B \epsilon^{\tl\mu-\tl\nu} A_{\tl\mu} \partial_- A_{\tl\nu} 
	\right)   \\ 
	& \quad + 
	\int d^3x~~ \bar{W}_\mu(\tfrac{i }{2 \pi}\epsilon^{\mu\nu\rho} \partial_\nu + {\sgn}(\kappa_B) v^2 g^{\mu \rho}) W_\rho 
	\\ &\quad +\int d^3x \left(Z_\mu\left(\tfrac{i\kappa_B}{4 \pi} \epsilon^{\mu\nu\rho}\partial_\nu + g^{\mu \rho} |\kappa_B| v^2 \right)Z_\rho \right) \\
	&\quad+ \frac{1}{2 \pi} \int d^3 x\ \epsilon^{\mu\nu\rho} \bar{W}_\rho \left(  A_\mu - Z_\mu \right)  W_\nu \ .
	\end{split}
\end{equation}
where the indices $\tilde{\mu},\ldots$ run over $+,3$ only. The first
line of the action \eqref{lcgauge} contains the quadratic kinetic
terms for the $SU(N_B-1)$ gauge fields $A_+$ and $A_3$.
\footnote{In this paper we have worked, for definiteness, with the case 
in which the original gauge group of the bosonic theory is $SU(N_B)$. Had we 
instead started with a $U(N_B)$ bosonic theory, the Lagrangian \eqref{lcgauge} 
would have continued to apply with the change that $A_\mu$ would be a $U(N_B-1)$ 
gauge field. All the leading large $N_B$ computations and results presented in the rest of this this paper 
would go through unmodified in this case (because $U(N_B-1)$ and $SU(N_B-1)$ results differ only
at subleading order in $\frac{1}{N_B}$).}
The second
line of \eqref{lcgauge} has the kinetic term for the $SU(N_B-1)$
fundamental field $W_\mu$ and its (antifundamental) complex conjugate
${\bar W}_\mu$. All gauge indices (which have been suppressed for
readability) in \eqref{lcgauge} are contracted, i.e. the gauge index
structure of all terms in this line is ${\bar W}^a W_a$. The third
line in \eqref{lcgauge} contains the kinetic term for the real,
$SU(N_B-1)$ neutral field $Z_\mu$. The last line in \eqref{lcgauge}
contains the only interaction terms present in the Lagrangian. The
first interaction term - which will play a crucial role in what
follows - has the gauge contraction structure
${\bar W}^a A_a{}^b W_b$.  This term completes the derivative in the
kinetic part of the $W$ action into an $SU(N_B-1)$ covariant
derivative.  It follows, in particular, that the current carried by
the $W$ bosons that couples to the $SU(N_B-1)$ gauge field $A$ is
given by
\begin{equation}\label{currentofw}
(J^\mu_A)_a^b= \frac{1}{2 \pi} \epsilon^{\mu\nu \rho} (W_\nu)_a ({\bar W}_\rho)^b\ .
\end{equation}
Finally the second interaction terms in the last line of
\eqref{lcgauge} has the gauge structure $Z {\bar W}^a W_a$. In
strictly formal analogy with the discussion above we may define the
`$Z$ boson current'
\begin{equation}\label{currentofz}
J^\mu_Z= -\frac{1}{2 \pi} \epsilon^{\mu\nu \rho} (W_\nu)_a ({\bar W}_\rho)^a\ .
\end{equation}
Note that 
\begin{equation}\label{rbc}
J_Z^\mu= -\text{Tr}\, J_A^\mu\ .
\end{equation}
Using these definitions the Lagrangian \eqref{lcgauge} can be rewritten as 
\begin{equation}\label{lcg}
	\begin{split}
	S_{\text{E}}	&=\frac{i}{4 \pi}\int d^3x\ \text{Tr}\left(\kappa_B \epsilon^{\tl\mu-\tl\nu} A_{\tl\mu} \partial_- A_{\tl\nu} 
	\right)   \\ 
	& \quad + 
	\int d^3x\ \bar{W}_\mu(\tfrac{i }{2 \pi}\epsilon^{\mu\nu\rho} \partial_\nu + {\sgn}(\kappa_B) v^2 g^{\mu \rho}) W_\rho 
	\\ &\quad +\int d^3x\ Z_\mu\left(\tfrac{i\kappa_B}{4 \pi} \epsilon^{\mu\nu\rho}\partial_\nu + g^{\mu \rho} |\kappa_B| v^2 \right)Z_\rho \\
	&\quad + \int d^3 x  \left( \text{Tr} (A_{\tl\mu} J^{\tl\mu}_A) + J_Z^\mu Z_\mu \right)\ .
	\end{split}
\end{equation}
We will find it convenient to work in Fourier space. Our conventions
for moving between real and Fourier space are given by
\begin{equation}
\begin{split}
\psi(x)=
\int \frac{\td^3 q}{(2 \pi)^3} \ e^{i x\cdot q} \psi(q) \ ,
\end{split}
\end{equation}
where $\psi(x)$ is any field and the measure $\td^3q$ is defined as
follows. In $\mbb{R}^3$ (i.e.~when we are interested in zero
temperature physics) the integration measure over momenta is usually
written as $d^3p$. While this measure is perfectly correct as written
at zero temperature, at finite temperature we are working on
$\mbb{R}^2 \times S^1$. In this space the measure along $\mbb{R}^2$ is
the usual $dp_1 dp_2$, but the measure for the momentum $p_3$ along
$S^1$ which we write as $\td p_3$ is different and is given by
\begin{equation}\label{wwmbddp}
  \int \mc{D}p_3\, f(p_3) = \int_{-\pi}^{\pi} \rho_B(\alpha) d\alpha\   
  \frac{2\pi}{\beta}  \sum_{n=-\infty}^\infty  f\left(\frac{2 \pi n + \alpha}{\beta}\right)\ ,
\end{equation}
where $\rho_B(\alpha)$ is the holonomy eigenvalue distribution defined
above and $f$ is any function of $p_3$. The action \eqref{lcg} may
then be rewritten as
\begin{equation}\label{lcgmom}
	\begin{split}
	S_{\text{E}} &= \int \frac{\td^3p}{(2\pi)^3}\ \text{Tr}\left( \frac{A_{\tilde{\mu}}(-p) K^{\tilde{\mu} \tilde{\rho}}(p)A_{\tilde{\rho}}(p)}{2} + J^{\tilde{\mu}}_A(-p) A_{\tilde{\mu}}(p) \right)  \\
	&+ \int \frac{\td^3p}{(2\pi)^3}\ \left( \frac{Z_\mu(-p)  K_Z^{\mu\rho}(p) Z_\rho(p)}{2} + J^\mu_Z(-p) Z_\mu(p) \right)\\
	&+\int \frac{\td^3p}{(2\pi)^3}\  {\bar W}_\mu(-p) K_W^{\mu\rho}(p) W_\rho(p)\ .
	\end{split}
\end{equation}
where
\begin{equation}\label{lfmd} \begin{split}
& K^{\tilde{\mu} \tilde{\rho}}(p) = \frac{-\kappa_B}{2 \pi} \epsilon^{\tilde{\mu}-\tilde{\rho}}p_-\ ,  \\
& K_Z^{\mu\rho}(p) = \frac{-\kappa_B}{4 \pi} \epsilon^{\mu\nu\rho}p_\nu + |\kappa_B| v^2 g^{\mu\rho}\ , \\
&K_W^{\mu\rho}(p) = -\frac{1}{2 \pi}
    \epsilon^{\mu\nu\rho} p_\nu + {\sgn}(\kappa_B) v^2
    g^{\mu\rho}\ , \\
& J_A^{\tilde{\mu}}(p) = \frac{1}{2 \pi} \epsilon^{\tilde{\mu}\nu\rho} \int \frac{d^3q}{(2\pi)^3} W_\nu(p-q) \bar{W}_\rho(q)\ , \\
& J_Z^\mu(p) = -\text{Tr}\, J_A^\mu(p)\ .
\end{split}
\end{equation}
A path integral based on the action \eqref{lcgmom} can be simplified
by integrating out the fields $A_\mu$ and $Z_\mu$.  As each of these
fields enter the action \eqref{lcgmom} quadratically, this integrating
out procedure can be performed exactly. For each of $A$ and $Z$ we
have to complete squares and evaluate a quadratic Gaussian
integral. Let us first ignore the determinants and simply classically 
eliminate the variables $A_\mu$ and $Z_\mu$ by completing 
squares. This procedure gives us an effective action for the 
$W_\mu$ and ${\bar W}_\mu$ fields given by 
\begin{equation}\label{asbm}
\begin{split}
&S_{\text{E}}[W] 
= \int \frac{\td^3p}{(2\pi)^3}\, \bar{W}_{a,\mu}(-p)\, K_W^{\mu\rho}(p)\, W^a_\rho(p) \\
& \quad \quad \quad \quad - \frac{1}{2} \int \frac{\td^3p}{(2\pi)^3}\frac{\td^3q}{(2\pi)^3}\frac{\td^3q'}{(2\pi)^3}\,  [\bar{W}_{\alpha} W_\beta](q,-p)\ \Lambda^{\alpha\beta\alpha'\beta'}(q-q',p)\  
[\bar{W}_{\alpha'} W_{\beta'}](q',p)\ ,
\end{split}
\end{equation}
with
\begin{equation}\label{asbm1}
\begin{split}
& K^{-1}_{\tilde{\mu}\tilde{\nu}}(p)  = \frac{2\pi}{\kappa_B p_-} \epsilon_{\tilde{\mu}-\tilde{\nu}}\ ,\  K^{-1}_{\tilde{\mu}\tilde{\nu}}(p) K^{ \tilde{\nu} \tilde{\rho}}(p)=\delta^{\tilde{\rho}}_{\tilde{\mu}} \\
&  K^{-1}_{Z,\mu\nu}(p) = \frac{-2\pi m_Z}{|\kappa_B| (p^2 + m_Z^2)} \left(\delta_{\mu\nu} - \text{sgn}(\kappa_B) \epsilon_{\mu\nu\rho} \frac{p^\rho}{m_Z} + \frac{p_\mu p_\nu}{m_Z^2}\right)\ ,\ K^{-1}_{Z,\mu\nu}(p) K_{Z}^{\nu \rho}(p) =\delta^{\rho}_{\mu} \\
& \Lambda^{\alpha\beta\alpha'\beta'}(q-q',p)  =
\Lambda_A^{\alpha\beta\alpha'\beta'}(q-q') +
\Lambda_Z^{\alpha\beta\alpha'\beta'}(p)\ ,\\
&\Lambda_A^{\alpha\beta\alpha'\beta'}(q-q') =   \frac{1}{(2\pi)^2} \epsilon^{\beta\alpha'\tilde{\mu}} K^{-1}_{\tilde{\mu}\tilde{\mu}'}(q-q') \epsilon^{\tilde{\mu}'\beta'\alpha}\ , \\
&\Lambda_Z^{\alpha\beta\alpha'\beta'}(p) =   \frac{1}{(2\pi)^2} \epsilon^{\alpha\beta\mu} K^{-1}_{Z,\mu\mu'}(p) \epsilon^{\mu'\alpha'\beta'}\ .\nonumber
\end{split}
\end{equation}
We have used the notation $[ B A]$ to denote the singlet combination
$B_a A^a$ where $B$ and $A$ are, respectively, fields that transform
in the antifundamental and fundamental of $SU(N_B-1)$. Moreover the
expression $[\bar{W}_{\alpha}W_{\beta}](q,p)$ in \eqref{asbm} is
shorthand for
\begin{equation}\label{Wbilin}
[\bar{W}_{\alpha}W_{\beta}](q,p) \equiv \left[\bar{W}_{\alpha}(q+\tfrac{p}{2}) W_{\beta}(-q+\tfrac{p}{2})\right]\ ,
\end{equation}
(where $p$ is the centre-of-mass momentum of the bilinear field
and $q$ its relative momentum).  We note that expressions can be
further simplified to give \footnote{Here we have used
  $\epsilon^{\beta \alpha'+}\epsilon^{3\beta' \alpha}=\epsilon^{\beta
    \alpha'+}( \epsilon^{3+-}
  \delta^{\alpha}_{-}\delta^{\beta'}_{+}+\epsilon^{3-+}
  \delta^{\alpha}_{+}\delta^{\beta'}_{-} ) $.}
\begin{multline}
\Lambda^{\mu\nu\mu'\nu'}(q-q',0) = \frac{1}{2\pi \kappa_B (q-q')_-} (\epsilon^{\nu \mu'\nu'} \delta_-^\mu - \epsilon^{\nu \mu'\mu} \delta_-^{\nu'}) \\ - \frac{1}{2\pi|\kappa_B| m_Z} (\delta^{\mu\mu'}\delta^{\nu\nu'} - \delta^{\mu\nu'} \delta^{\nu\mu'})\ . 
\end{multline}
For use in the next section we note the some easily verified symmetry properties of the quartic couplings $\Lambda_A$ and $\Lambda_Z$ above:
\begin{equation} \label{symp}
\begin{split}
& \Lambda_A^{\mu\nu \mu'\nu'}(p) = -\Lambda_A^{\mu\nu \mu'\nu'}(-p) = -\Lambda_A^{\mu\mu' \nu\nu'}(p) = -\Lambda^{\nu'\nu \mu'\mu}(p)\ ,\\
& \Lambda_Z^{\mu\nu \mu'\nu'}(0) = -\Lambda_Z^{\nu\mu \mu'\nu'}(0) = -\Lambda_Z^{\mu\nu \nu'\mu'}(0)\ .
\end{split}
\end{equation}
The final path integral we need to perform is given by 
\begin{equation}
\label{finpiin} 
\mathcal{Z} = \int [dW]  e^{-S_{\text{E}}[W]} \ \detr_{\!A} \ \detr_{\!Z}\ ,
\end{equation}
where $S_{\text{E}}[W]$ is the Euclidean action  listed in \eqref{asbm}, 
$\detr_{\!A}$ is the determinant that results from integrating out the 
$A_\mu$ fields and $\detr_{\!Z}$ is the determinant resulting from 
integrating out the $Z_\mu$ fields. We now turn to a study of 
these two determinants. 

It is easily verified that 
\begin{equation}\label{Adet}
\detr_{\!A} = e^{-\int \frac{\td^3p}{(2\pi)^3} \log \frac{i\kappa_B p_-}{4\pi}}
\end{equation}
This determinant is formally cancelled by the Faddeev Popov determinant
associated with the gauge fixing to $A_-=0$ and so may be
discarded\footnote{At any event this determinant and its Faddeev-Popov
  counterpart are both independent of temperature, the $W$ fields and
  the holonomy fields and so can be absorbed into the normalization of
  the path integral (equivalently into a shift of the ground state
  energy) and so can be ignored.}.  On the other hand we find
\begin{equation}
\label{Zdet}
\detr_{\!Z} = e^{-\frac{1}{2}\int \frac{\td^3p}{(2\pi)^3} \log \text{det} K_Z^{\nu\rho}(p)} =  e^{-\int \frac{\mc{V}_2 d^2 p_s}{(2\pi)^2}  \log\left( 1-e^{\beta \sqrt{p_s^2 +4 \pi v^2 } }\right) }\ ,
\end{equation}
where $\mc{V}_2$ is the volume of two dimensional space. \eqref{Zdet}
is a nontrivial function of temperature, but contributes to the
logarithm of the partition function only at order unity. The
contribution we will obtain below from integrating over the $N_B$ W
bosons will clearly be of order $N_B$. Consequently the contribution
of \eqref{Zdet} to the free energy is subleading in an expansion in
$\frac{1}{N_B}$ and we ignore it in what follows.

In summary, at leading order in the large $N_B$ limit we can simply
ignore both determinants $\detr_A$ and $\detr_Z$ and work with the
simplified path integral
\begin{equation}
\label{finpi} 
\mc{Z} = \int [dW] e^{-S_{\text{E}}[W]}\ ,
\end{equation}
with $S_{\text{E}}[W]$ given in \eqref{asbm}.  We now proceed to evaluate this
path integral in the large $N_B$ limit.

\subsection{Dynamics in terms of singlet fields}
In order to exploit the simplifications of the large $N_B$ limit we
imitate the analysis of \cite{Jain:2012qi} and employ a variant of the Hubbard-Stratonovich trick. Specifically we introduce two bilocal but
$SU(N_B-1)$ singlet auxiliary fields
$\Sigma^{\mu\nu}(q,p)$ and $\alpha_{\mu\nu}(q,p)$ and introduce
these into the path integral using the identities
\begin{equation} \label{alphsig}
	\begin{split}
		1 &= \int [d\alpha] \  \delta\left[ \kappa_B \alpha_{\mu\nu}(q,p) + [\bar{W}_\mu W_\nu](q,p) \right] \\ & =\int [d\alpha][d\Sigma]\ \exp\left(\int \frac{\td^3p}{(2\pi)^3}\frac{\td^3q}{(2\pi)^3}\ i \Sigma^{\nu\mu}(-q,-p) \left(\kappa_B \alpha_{\mu\nu}(q,p) + [\bar{W}_\mu W_\nu](q,p) \right)\right)
	\end{split}
\end{equation}
Recall the definition of $[\bar{W}_\mu W_\nu](q,p)$ from
\eqref{Wbilin}. Similarly, $p$ is to be thought of as the
centre-of-mass momentum and $q$ the relative momentum of the bilocal
fields $\alpha_{\mu\nu}(q,p)$ and $\Sigma^{\mu\nu}(q,p)$.
% Below we work with the Fourier transform of the bilocal fields 
% $\alpha$ and $\Sigma$ defined by 
% \begin{equation} \label{FTA}
% \begin{split}
% &\alpha_{\mu\nu}(x, y)=\int 
% \frac{d^3q}{(2 \pi)^3}\frac{d^3p}{(2 \pi)^3} \ e^{iq\cdot (x-y)}e^{ip\cdot \frac{x+y}{2}}\, \alpha_{\mu\nu}(q, p) \\
% &\Sigma^{\mu\nu}(x, y)=\int 
% \frac{d^3q}{(2 \pi)^3}\frac{d^3p}{(2 \pi)^3} \ e^{iq\cdot (x-y)}e^{ip\cdot  \frac{x+y}{2}}\, \Sigma^{\mu\nu}(q, p) 
% \end{split}
% \end{equation}
Inserting the identity \eqref{alphsig} into the path integral,
the action \eqref{asb} can be written as
\begin{equation} \label{naa}
	\begin{split}
		&\frac{S_{E}[\alpha,\Sigma,W]}{N_B}=  -\frac{i}{\lambda_B} \ \int \frac{\td^3p}{(2 \pi)^3}\frac{\td^3q}{(2 \pi)^3}  
	\Sigma^{\nu\mu}(q, p) \alpha_{\mu\nu}(-q,-p) \\
	& +\frac{1}{N_B} \int \frac{\td^3q}{(2 \pi)^3}\frac{\td^3p}{(2 \pi)^3} \bar{W}_\mu(-q-\tfrac{p}{2})\,Q^{\mu\nu}(q,p)\, W_\nu (q-\tfrac{p}{2}) \\
	&- \frac{1}{2\lambda_B}\int \frac{\td^3p}{(2\pi)^3}\frac{\td^3q}{(2\pi)^3}\frac{\td^3q'}{(2\pi)^3}\,  \alpha_{\mu\nu}(q,-p)\ \kappa_B\Lambda^{\mu\nu\mu'\nu'}(q-q',p)\  
  \alpha_{\mu'\nu'}(q',p)\ .
		\end{split}
\end{equation}
where 
\begin{align}\label{qdef}
Q^{\mu\nu}(q,p) &= (2 \pi)^3 \delta(p) K_W^{\mu\nu}(q)  
  -i \Sigma^{\nu\mu}(q, p)\ ,\nonumber\\
\text{or}\quad Q (q,p) &= (2\pi)^3 \delta(p) K_W(q) - i \Sigma^T(q,p)\quad\text{schematically}\ .  
\end{align}
and the quartic coupling $\Lambda$ is defined in \eqref{asbm}. It is useful to define
\begin{equation}\label{Vdef}
V[\alpha] = -\frac{1}{2\lambda_B}\int \frac{\td^3p}{(2\pi)^3}\frac{\td^3q}{(2\pi)^3}\frac{\td^3q'}{(2\pi)^3}\,  \alpha_{\mu\nu}(q,-p)\ \kappa_B\Lambda^{\mu\nu\mu'\nu'}(q-q',p)\  
  \alpha_{\mu'\nu'}(q',p)\ .
\end{equation}
% \begin{equation} \label{eaal}
% \begin{split}
% &V=- \frac{i }{2 \pi \lambda_B} 
% \int  \int \frac{d^3p}{(2 \pi)^3} \frac{d^3q_1}{(2 \pi)^3}  \frac{d^3q_2}{(2 \pi)^3} 
% \frac{1}{(q_1-q_2)_{-}} \\
% & \bigg[ \ \alpha_{--}(q_1 + \frac{p}{2} )\alpha_{3+}(q_2 - \frac{p}{2})-\alpha_{-3}(q_1+  \frac{p}{2})\alpha_{-+}(q_2 - \frac{p}{2}) \\ &+\alpha_{+3}(q_1 + \frac{p}{2})\alpha_{--}(q_2- \frac{p}{2}) -\alpha_{+-}(q_1  +\frac{p}{2})\alpha_{3-}(q_2 - \frac{p}{2} )   \  \bigg]
% \end{split}
% \end{equation}
in terms of which the effective action takes the form
\begin{equation} \label{naan}
	\begin{split}
		&\frac{S_{E}[\alpha,\Sigma,W]}{N_B}= \frac{1}{N_B} \int \frac{\td^3q}{(2 \pi)^3}\frac{\td^3p}{(2 \pi)^3} 
		\bar{W}_\mu(-q-\tfrac{p}{2})\,Q^{\mu\nu}(q,p)\, W_\nu (q-\tfrac{p}{2}) \\
		& \quad \quad \quad \quad \quad \quad \quad + V[\alpha] -\frac{i}{\lambda_B} \ \int \frac{\td^3p}{(2 \pi)^3}\frac{\td^3q}{(2 \pi)^3}  
	\Sigma^{\nu\mu}(q, p) \alpha_{\mu\nu}(-q,-p) .\\
		\end{split}
\end{equation}
As the effective action \eqref{naan} is a quadratic function of the 
$W$-bosons, they can be integrated out. The result of such an
integration is an effective action for the bilocal fields that takes
the schematic form
\begin{equation}\label{ema}
\begin{split}
S_{\text{eff}}[\alpha, \Sigma] = N_B\left(-\frac{i}{\lambda_B} \ \Sigma\cdot\alpha + \log \detr Q + V[\alpha]  \right)\ .
\end{split}
\end{equation}
As the action \eqref{ema} is of order $N_B$ the subsequent integral
over the $\Sigma$ and $\alpha$ fields can be performed  - at leading order in 
$1/ N_B$ - in the saddle point approximation. We will assume
that the saddle point solution for $\Sigma$ and $\alpha$ preserves
translational invariance, i.e. that the saddle point solution takes
the form
\begin{equation} \label{sps} \begin{split}
	&\Sigma^{\mu\nu}(q, p) = (2\pi)^3 \delta(p) \Sigma^{\mu\nu}(q)\ , \\
	 &\alpha_{\mu\nu}(q, p) = (2\pi)^3 \delta(p) \alpha_{\mu\nu}(q)\ . 
	 \end{split}
\end{equation}
% \footnote{From here on till the end of this paper, the
% 	the field $\Sigma^{\mu\nu}$ and $\alpha_{\mu\nu}$  will always
% 	refer to the `single momentum fields that appear on the RHS rather 
% 	than the double momentum fields that appear on the LHS of \eqref{sps}}
Under this assumption the expression for $Q^{\mu\nu}(q,p)$ in
\eqref{qdef} simplifies to
\begin{equation}\label{qdefa}
\begin{split}
&Q(q,p)= (2 \pi)^3 \delta(p) Q(q)\ ,\quad\text{with}\quad Q(q)=K_W(q)  -i \Sigma^{T}(q)\ .
\end{split}
\end{equation}
From this point on every occurrence of the symbols $\Sigma^{\mu\nu}$,
$\alpha_{\mu\nu}$ and $Q^{\mu\nu}$ in this paper will refer to the
`single momentum' fields on the RHS of \eqref{sps} and \eqref{qdefa}
rather than the bi-momentum field on the LHS of \eqref{sps} and \eqref{qdefa}.

The integral over the W bosons in \eqref{naa} is now easily performed
and gives rise to the following effective action for the
$\alpha_{\mu\nu}$ and $\Sigma^{\mu\nu}$ fields:
\begin{equation} \label{naab}
  \frac{S_{\text{eff}}[\alpha,\Sigma]}{N_B \mathcal{V}_3} =
  V_0[\alpha] -\frac{i}{\lambda_B} \ \int \frac{\td^3q}{(2 \pi)^3}
  \Sigma^{\nu\mu}(q) \alpha_{\mu\nu}(-q) + \int \frac{\td^3q}{(2 \pi)^3}
  \log {\rm det} \left(K_W(q) -i \Sigma^{T}(q) \right) \ .
\end{equation}
where $\mathcal{V}_3 = \mathcal{V}_2 \beta$ is the volume of
spacetime. The quantity $V_0[\alpha]$ is obtained by setting the
centre-of-mass momentum $p$ to zero in the integrand of $V[\alpha]$ in
\eqref{Vdef} and dividing by $\mc{V}_3$:
\begin{align}
&V_0[\alpha] = -\frac{1}{2\lambda_B}\int \frac{\td^3q}{(2\pi)^3}\frac{\td^3q'}{(2\pi)^3}\,  \alpha_{\mu\nu}(q)\ \kappa_B \Lambda^{\mu\nu\mu'\nu'}(q-q',0)\  
\alpha_{\mu'\nu'}(q')\ 
\end{align}

\subsection{A symmetry of the gap equations}
Varying the action \eqref{naab} w.r.t $\Sigma^{\mu\nu}(-q)$ yields the 
equation
\begin{equation}
\label{wia}
\alpha_{\nu\mu}(q) = \lambda_B \frac{\delta}{i\delta \Sigma^{\mu \nu}(-q)} \log {\rm det}
\left(K_W(q)  -i \Sigma^{T}(q) \right)\ .
\end{equation}
We might have anticipated from \eqref{alphsig} that the on-shell value
of $\alpha_{\mu\nu}$ would turn out to be the (appropriately normalized
and colour stripped) propagator of the $W$ bosons, while
$\Sigma^{\mu\nu}$ would turn out to be the self energy in this
propagator. This expectation is confirmed by the explicit form of
\eqref{wia}.

Varying \eqref{naab} w.r.t. the $\alpha_{\mu\nu}(-q)$ yields an expression 
for the self energy $\Sigma^{\mu\nu}$ in terms of $\alpha_{\mu\nu}$ 
and so - using \eqref{wia} - in terms of $\Sigma^{\mu\nu}$. Explicitly 
\begin{equation}\label{sig}
\Sigma^{\nu\mu}(q) = \frac{i}{2} \int \frac{\td^3q'}{(2\pi)^3} \left(\kappa_B \Lambda^{\mu\nu\mu'\nu'}(q'-q,0) + \kappa_B \Lambda^{\mu'\nu'\mu\nu}(q-q',0)\right) \alpha_{\mu'\nu'}(-q')\ .
\end{equation}

Before turning to the structure of the RHS of \eqref{sig} in detail,
we pause to note an important symmetry property of solutions to this equation. 
Using the explicit expressions in \eqref{lfmd} we can immediately 
verify that 
\begin{equation}
\label{kw}
K_W^{\mu\nu}(q)= K_W^{\nu\mu}(-q)\ ,\quad\text{i.e.}\quad K_W(q) = K_W^T(-q)\ .
\end{equation}
It follows that, at tree level, 
\begin{equation}
\label{kwb}
\alpha(q)_{\mu \nu}= \alpha_{\nu \mu}(-q)\ ,\quad\text{i.e.}\quad \alpha(q) = \alpha^T(-q)\ 
\end{equation}

We will now demonstrate that relations analogous to \eqref{kw} 
and \eqref{kwb} apply not just at tree level but at every order 
in perturbation theory. 

Let us first work at the lowest nontrivial order in perturbation
theory. In order to obtain the `one loop' contribution to the $W$
boson self energy we plug the tree level propagator $\alpha_{\mu\nu}$
(which is of order $\lambda_B$) into the RHS of the equation
\eqref{sigmafull}
\begin{align}\label{sigmafull}
\Sigma^{\nu\mu}(q) &= \frac{i}{4\pi} \int \frac{\td^3q'}{(2\pi)^3} \left(\epsilon^{\nu \mu'\nu'} \delta_-^\mu - \epsilon^{\nu \mu'\mu} \delta_-^{\nu'} - \epsilon^{\nu' \mu\nu} \delta_-^{\mu'} + \epsilon^{\nu' \mu\mu'} \delta_-^{\nu}\right) \frac{\alpha_{\mu'\nu'}(-q')}{(q'-q)_-} + \nonumber\\ 
&\quad -\frac{i \text{sgn}(\kappa_B)}{2\pi m_Z} \int  \frac{\td^3q'}{(2\pi)^3} (\delta^{\mu\mu'}\delta^{\nu\nu'} - \delta^{\mu\nu'}\delta^{\nu\mu'}) \alpha_{\mu'\nu'}(-q')\ .
\end{align}
The resultant expression is of order $\lambda_B$ and is the first
order (or one loop) correction to $\Sigma^{\mu\nu}$.  Using
\eqref{kwb}, it is easy to verify that the second line of
\eqref{sigmafull} vanishes. The first line of \eqref{sigmafull} does not
vanish and gives a nonzero one loop contribution to
$\Sigma^{\mu\nu}$. Using \eqref{kwb} however, it is easy to convince
oneself that this first order correction to $\Sigma^{\mu\nu}$ obeys
\begin{equation}
\label{kwa}
\Sigma^{\mu\nu}(q)= \Sigma^{\nu\mu}(-q)\ ,\quad\text{i.e.}\quad \Sigma(q) = \Sigma^T(-q)\ .
\end{equation}
It follows that up to first order in $\lambda_B$ 
\begin{equation}
\label{kwq}
Q^{\mu\nu}(q)= Q^{\nu\mu}(-q)\ ,\quad\text{i.e.}\quad Q(q) = Q^T(-q)\ .
\end{equation}
which implies that $\alpha_{\mu\nu}$ obeys \eqref{kwb} up to first subleading order in $\lambda_B$. 

This argument can now be iterated. In order to obtain the `two loop'
contribution to $\Sigma^{\mu\nu}$ one plugs the
${\cal O}(\lambda_B^2)$ part of $\alpha_{\mu\nu}$ into the RHS of
\eqref{sigmafull} and evaluates the integrals on the RHS. The fact
that this correction piece in $\alpha_{\mu\nu}$ also obeys \eqref{kwb}
implies the two loop correction to $\Sigma^{\mu\nu}$ receives no
contribution from the second line of \eqref{sigmafull}.  The entire
contribution to this two loop correction comes from the first line of
\eqref{sigmafull}, which, in turn, now obeys \eqref{kwa}. It follows
that $Q^{\mu\nu}$ obeys \eqref{kwq} to second order. This implies
$\alpha_{\mu\nu}$ obeys \eqref{kwb} upto first subleading order in
$\lambda_B$ and so on. Iterating the argument above indefinitely we
conclude
\begin{itemize}
\item The equations \eqref{kwq}, \eqref{kwb} and \eqref{kwa} are obeyed at every order 
in the $\lambda_B$ expansion.
\item The contribution of  second line in \eqref{sigmafull} vanishes at 
every order in the $\lambda_B$ expansion. 
\end{itemize}
Note that the second line in \eqref{sigmafull} summarizes the
contribution of $Z_\mu$ exchange to $\Sigma^{\mu\nu}$. The fact that
this line does not contribute to the gap equation at any order in
$\lambda_B$, and so can just be dropped, tells us that that diagrams
involving propagating $Z_\mu$ bosons do not contribute to the
partition function at leading order in the large $N_B$ limit. Note
that this conclusion does not follow from large $N_B$ counting, but
instead follows from the slightly more detailed analysis presented
above \footnote{At the diagrammatic level this is the assertion that
  `tadpole' contributions to $\Sigma^{\mu\nu}$ - the second graph on
  the RHS of Fig. \ref{fii} vanishes at all orders.}.

In summary, our saddle-point equations or \emph{gap equations} take
the final form\footnote{We have changed variables from $-q'$ to $q'$
  in going from \eqref{sigmafull} to \eqref{sigmagap}. To get the second equality in \eqref{alphagap} we have used that for any non-singular matrix $M$ we have $\log \detr M=\Tr \log M \implies \delta(\log \detr M)= \Tr (M^{-1} \delta M)$.}:
\begin{align} 
\alpha_{\nu\mu}(q) &= \lambda_B \frac{\delta}{i\delta \Sigma^{\mu \nu}(-q)} \log {\rm det} \left(K_W(q)  -i \Sigma^{T}(q) \right) = - \lambda_B (Q^{-1}(q))_{\nu\mu}\ ,
 \label{alphagap} \\
  \Sigma^{\nu\mu}(q) &= -\frac{i}{4\pi} \int \frac{\td^3q'}{(2\pi)^3} \left(\epsilon^{\nu \mu'\nu'} \delta_-^\mu - \epsilon^{\nu \mu'\mu} \delta_-^{\nu'} - \epsilon^{\nu' \mu\nu} \delta_-^{\mu'} + \epsilon^{\nu' \mu\mu'} \delta_-^{\nu}\right) \frac{\alpha_{\mu'\nu'}(q')}{(q'+q)_-}\ .\label{sigmagap}
\end{align}
where $\Sigma^{\mu\nu}$, $\alpha_{\mu\nu}$ and $Q^{\mu\nu}$ enjoy the symmetry
properties \eqref{kwa}, \eqref{kwa} and \eqref{kwq} respectively. 
%\footnote{The first equation in \eqref{sigmagap} is equivalent to  \eqref{wia} given the symmetry properties \eqref{kw} and \eqref{kwa}.} 

Our final gap equation, \eqref{sigmagap}, may be diagrammatically
summarized as in Fig. \ref{fii}. The LHS of the figure is the $W$
boson self energy. On the RHS of the figure the double lines denote
the exact $W$ boson propagators, the dashed line is the $Z$ boson
propagator while the wiggly line is the $SU(N_B-1)$ gauge boson
propagator. The RHS of \eqref{sigmagap} is entirely captured by the
first figure on the RHS of Fig. \ref{fii}. This is consistent because
the second figure on the RHS of Fig \ref{fii} (i.e. the contribution
to $\Sigma$ of $Z$ boson exchange) vanishes, as we have demonstrated
above.

\newsavebox{\feynrules}
\sbox{\feynrules}{
\begin{fmffile}{vertices} % file name and path
  \fmfset{thin}{.8pt}
  \fmfset{wiggly_len}{2mm}
  \fmfset{dash_len}{2.5mm}
  \fmfset{dot_size}{1thick}
  \fmfset{arrow_len}{2.5mm}
  \fmfset{curly_len}{2.5mm}
  \fmfset{unitlength}{2mm}
  
  \begin{fmfgraph*}(50,50)
    \fmfkeep{sew}
    \fmfleft{i} \fmfright{o}
%    \fmf{double,label=$p$}{i,v1}
    \fmf{heavy,tension=1.2,label=$q'$}{i,o}
 %   \fmf{double,label=$p$}{v2,o}
    \fmfdot{i,o}
    \fmf{photon,left,tension=0.5,label=$q-q'$}{i,o}
  \end{fmfgraph*}

  \begin{fmfgraph*}(50,50)
    \fmfkeep{sez}
    \fmfbottom{i} \fmftop{o}
    \fmf{dashes,tension=0.5,label.left=$0$}{i,v1}
    \fmf{heavy,left,tension=0.3,label=$q'$}{v1,v2}
    \fmf{double,left,tension=0.3}{v2,v1}
    \fmf{phantom}{v2,o}
%    \fmf{double,tension=0.5}{v1,o}
    \fmfdot{i}
%    \fmf{photon,left,tension=0,label=$p-q$}{v1,v2}
  \end{fmfgraph*}
\end{fmffile}
}
%\vspace{10pt}
\begin{equation}
\label{fii}
\Sigma(q)\quad =\quad \fmfvcenter{sew}\quad + \quad \text{\raisebox{5ex}{$\fmfvcenter{sez}$}}
\end{equation}

\subsection{Reduction to integral equations of a single variable}

Notice that the RHS of the gap equation \eqref{sigmagap} or
\eqref{sigmacomp} for $\Sigma(q)$ is independent of $q_3$.  It follows
that each component of  $\Sigma^{\mu\nu}$ is independent of $q_3$. 

The various components of $\Sigma^{\mu\nu}$ can, in general, depend on
$q^+$ and $q^-$. To further constrain this dependence we note that our
gauge choice $A_-=0$ preserves an $SO(2)$ subgroup of the Euclidean
isometry group $SO(3)$ of our theory. We choose conventions so that
$q_\pm$ carries unit positive(negative) charge under this subgroup;
the general rule is that every lower $+$ and upper $-$ sign carries
positive unit charge, while every lower $-$ and upper $+$ sign carries
negative unit charge. It is easy to verify that, with these
conventions, $SO(2)$ rotations are a symmetry of the gap equations.
It follows that any given component of $\Sigma^{\mu\nu}$ must be given
by a number of explicit powers of $q_{\pm}$ (determined by the charge
of that component) times an unknown function of
\begin{equation}\label{qssq}
w = q_s^2 = 2q_+q_-\ .
\end{equation}
More specifically, we choose to parametrize  non-zero components of $\Sigma^{\mu\nu}$ as 
\begin{equation}
\label{Fdefinition}
\begin{split}
\Sigma^{--}(q) &= \frac{1}{2\pi q_-^2} F_1(w)\ ,\\
\Sigma^{+-}(q) &= +\Sigma^{-+}(q) = \frac{1}{2\pi} F_2(w)\ ,\\
\Sigma^{3-}(q) &= -\Sigma^{-3}(q) = \frac{1}{2\pi q_-} F_3(w)\ ,\\
\Sigma^{3+}(q) &= -\Sigma^{+3}(q) = \frac{q_-}{2\pi} F_4(w)\ .
\end{split}
\end{equation}
We will now recast the gap equations as equations for  four unknown
single variable functions $F_1,\ldots,F_4$.

To start with it is useful to express the matrix $Q^{\mu\nu}$
explicitly in terms of the the functions $F_1,\ldots, F_4$.  We define
the quantity $m$ as
\begin{equation}\label{convn1}
m = -\frac{\lambda_B m_B^{\text{cri}}}{2} = \sgn(\kappa_B) 2\pi v^2\ ,
\end{equation}
where we have used \eqref{defv} to get the second equality above (see
also \eqref{convn} of Appendix \ref{review}). Note that $|m|=|m_W|$
(see \eqref{baremass}) and also the fact that $m$ changes sign as
$\kappa_B$ changes sign. With this notation we find the quadratic
kernel $Q^{\mu\nu}(q)$ as defined in \eqref{qdefa} (the matrix is
presented in the order $+$, $-$, $3$)
\begin{equation}   \label{qitfoff}
Q^{\mu\nu}(q)=\frac{1}{2\pi}\begin{bmatrix}
0  & -i( F_2 + im + q_3)   & i q_-(1 - F_{4})  \\
-i( F_{2} + im - q_3) &  -\frac{i}{q_-^2} F_{1}(w) & -\frac{i}{q_-}(F_3 + \tfrac{w}{2})\\
-i q_-(1 - F_{4}) & \frac{i}{q_-}(F_{3} + \tfrac{w}{2}) &  m
\end{bmatrix}\ .
\end{equation}
The zeros of determinant of the matrix $Q^{\mu\nu}$ are the pole mass
of the propagator. The determinant is given by
\begin{equation}\label{detQ}
\begin{split}
&\detr Q =  -\frac{m}{8\pi^3} (q^2 + M^2(w))\ \\
& q^2 = w + q_3^2\\
& M^2(w) = -(F_2+im)^2 - \tfrac{i}{m} F_1 (1 - F_4)^2 - \tfrac{i}{m}(F_2+im)(w + 2 F_3)(1 - F_4) -  w\ .
\end{split} 
\end{equation}
With these formulae at hand now we proceed to give explicit form of
the gap equations for $\Sigma, \alpha$.

The gap equation \eqref{sigmagap} is given by
\begin{equation}\label{sigmacomp}
\begin{split}
& \Sigma^{33}(q)=0\ ,\\
& \Sigma^{++}(q)=0\ ,\\
&\Sigma^{--}(q)=-\frac{1}{ \pi}\int \frac{\td^3q'}{(2 \pi)^3}\frac{1}{(q+q')_{-}}\alpha_{3+}(q')\ ,\\
& \Sigma^{+-}(q)=-\frac{1}{2 \pi}\int \frac{\td^3q'}{(2 \pi)^3}\frac{1}{(q+q')_{-}}\alpha_{-3}(q')\ ,\\
& \Sigma^{3-}(q)=\frac{1}{2 \pi}\int \frac{\td^3q'}{(2 \pi)^3}\frac{1}{(q+q')_{-}}\alpha_{-+}(q')\ ,\\
& \Sigma^{3+}(q)=-\frac{1}{2 \pi}\int \frac{\td^3q'}{(2 \pi)^3}\frac{1}{(q+q')_{-}}\alpha_{--}(q')\ .
\end{split}
\end{equation}
It follows from \eqref{alphagap}) that the 
components of $\alpha(q) = -\lambda_B Q(q)^{-1}$ are
\begin{align}\label{alphaexpr}
  \alpha_{++}(q) &= \frac{\lambda_B}{(2\pi)^2\detr Q}\, \frac{1}{q_-^2}\left(imF_1 + (F_3 + \tfrac{w}{2})^2\right)\ ,\nonumber\\
  \alpha_{-+}(q) &= \frac{\lambda_B}{(2\pi)^2\detr Q}\left((1 - F_4)(F_3+\tfrac{w}{2}) - im( F_2 + im - q_3)\right)\ ,\nonumber\\
  \alpha_{--}(q) &=  \frac{\lambda_B}{(2\pi)^2\detr Q}\, q_-^2 (1 - F_4)^2\ ,\nonumber\\
  \alpha_{-3}(q) &=  -\frac{\lambda_B}{(2\pi)^2\detr Q}\, q_- (1 - F_4) (F_2 + i m - q_3)\ ,\nonumber\\
  \alpha_{3+}(q) &=- \frac{\lambda_B}{(2\pi)^2 \detr Q} \frac{1}{q_-}\left( F_1 (1-F_4) + (F_2 + i m - q_3) (F_3 + \tfrac{w}{2})\right)\ ,\nonumber\\
    \alpha_{33}(q) &= -\frac{\lambda_B}{(2\pi)^2\detr Q} \left((F_2 + im)^2 - q_3^2\right)\ .
\end{align}
Inserting \eqref{alphaexpr} into \eqref{sigmagap} we find the 
explicit coupled integral equations:
\begin{align} \label{inteqexpi}
  \frac{1}{q_-^2} F_1(w) &= \frac{\lambda_B}{(2\pi)^2} \int \frac{\td^3 q'}{(2\pi)^3} \frac{2F_1 (1 - F_4) + (F_2 + im)(2 F_3 + w')}{ \detr Q(q')(q+q')_- q'_-}\ ,\nonumber\\
 F_2(w) &= \frac{\lambda_B}{(2\pi)^2}\int \frac{\td^3 q'}{(2\pi)^3} \frac{(1 - F_4) (F_2 + im - q'_3) q'_-}{\detr Q(q')(q+q')_-}\ ,\nonumber\\
  \frac{1}{q_-} F_3(w) &= \frac{\lambda_B}{(2\pi)^2} \int \frac{\td^3 q'}{(2\pi)^3} \frac{(F_3 + \tfrac{w'}{2})(1 - F_4) - im(F_2 + im - q'_3)}{ \detr Q(q')(q+q')_-}\ ,\nonumber\\
 q_- F_4(w) &= -\frac{\lambda_B}{(2\pi)^2}\int \frac{\td^3 q'}{(2\pi)^3} \frac{(1 - F_4)^2 q^{\prime 2}_-}{ \detr Q(q')(q+q')_- }\ .
\end{align}
(all the functions $F_1$, $F_2$, $F_3$ and $F_4$ on the RHS of
\eqref{inteqexpi} and have the argument $w'$). Substituting the
expression for $\detr Q$ from \eqref{detQ} we obtain
\begin{align} \label{inteqexp}
  \frac{1}{q_-^2} F_1(w) &= -\frac{ 2 \pi \lambda_B}{m} \int \frac{\td^3 q'}{(2\pi)^3} \frac{2F_1 (1 - F_4) + (F_2 + im)(2 F_3 + w')}{  \left( (q_3')^2 +w' + M^2(w') \right)  ((q+q')_- q'_-)}\ ,\nonumber\\
 F_2(w) &= -\frac{ 2 \pi \lambda_B}{m}\int \frac{\td^3 q'}{(2\pi)^3} \frac{(1 - F_4) (F_2 + im - q'_3) q'_-}{\left( (q_3')^2 +w' + M^2(w') \right) (q+q')_-}\ , \nonumber\\
  \frac{1}{q_-} F_3(w) &= -\frac{2 \pi \lambda_B}{m} \int \frac{\td^3 q'}{(2\pi)^3} \frac{(F_3 + \tfrac{w'}{2})(1 - F_4) - im(F_2 + im - q'_3)}{ \left( (q_3')^2 +w' + M^2(w') \right) (q+q')_-}\ ,\nonumber\\
 q_- F_4(w) &= \frac{2 \pi \lambda_B}{m}\int \frac{\td^3 q'}{(2\pi)^3} \frac{(1 - F_4)^2 q^{\prime 2}_-}{ 
 \left( (q_3')^2 +w' + M^2(w') \right) (q+q')_- }\ .
\end{align}
The dependence of the integrands on the RHS of \eqref{inteqexp} on
$q'_3$ is completely explicit since the unknown functions on the RHS
are all functions of $w'$. As a consequence the integral (sum) over
$q_3'$ is easily evaluated as we now demonstrate. Recall, from
equation \eqref{wwmbddp}, that the `integral over $q_3$' is really a
discrete sum
\begin{equation}
 \int \frac{\mc{D}q_3}{2\pi} f(q_3) = \int d\alpha \rho_B(\alpha) \beta^{-1}\sum_{n\,\in\,\mathbb{Z}} f\left( \frac{\alpha+ 2\pi n }{\beta}\right)\ .
\end{equation}
To proceed, we make the assumption that the holonomy distribution
$\rho_B(\alpha)$ is an even function of $\alpha$:
\begin{equation}\label{rhoeven}
  \rho_B(\alpha) = \rho_B(-\alpha)\ .
\end{equation}
It follows that all integrals with an odd number of $q'_3$ factors
vanish in \eqref{inteqexp}. In terms of the function $\chi(z)$ defined
as
\begin{align}\label{intdet}
  \chi(z) &\equiv -\frac{(2\pi)^3}{m \beta} \int d\alpha \rho_B(\alpha)\, \sum_{n\,\in\,\mathbb{Z}} \frac{1}{(2 \pi \frac{n}{\beta}+\frac{\alpha}{\beta})^2+(z+M^2(z))}\ ,\\
    %      &= \frac{2\pi\beta}{m} \int d\alpha \rho_B(\alpha)\sum_{n\,\in\,\mathbb{Z}} \frac{1}{( n + \tfrac{\beta}{2\pi}\frac{\alpha}{\beta})^2+ \tfrac{\beta^2}{4\pi^2}(z+M^2(z))}\ ,\nonumber\\
          &= -\frac{2\pi^3}{m} \int d\alpha \rho_B(\alpha) \frac{1}{\sqrt{z + M^2(z)}}\times \nonumber \\ &\qquad\qquad\qquad \times \left(\coth (\tfrac{\beta}{2}(\sqrt{z + M^2(z)} + i\tfrac{\alpha}{\beta})) + \coth (\tfrac{\beta}{2}(\sqrt{z + M^2(z)} - i\tfrac{\alpha}{\beta}))\right)\ ,\nonumber
\end{align}
the integral equations \eqref{inteqexp} become
\begin{align} \label{inteqexp2}
  \frac{1}{q_-^2} F_1(w) &= \frac{\lambda_B}{(2\pi)^2}\int \frac{d^2 q'}{(2\pi)^2} \chi(w')\frac{2F_1(1-F_4) + (F_2 + im)(2 F_3 + w')}{(q+q')_- q'_-\,}\ ,\nonumber\\
  F_2(w) &= \frac{\lambda_B}{(2\pi)^2}\int \frac{d^2 q'}{(2\pi)^2} \chi(w')\frac{(1 - F_4) (F_2 + im) q'_-}{(q+q')_-}\ ,\nonumber\\
  \frac{1}{q_-} F_3(w) &= \frac{\lambda_B}{(2\pi)^2} \int \frac{d^2 q'}{(2\pi)^2} \chi(w')\frac{(F_3 + \tfrac{w'}{2})(1 - F_4) - im( F_2 + im)}{(q+q')_-}\ ,\nonumber\\
  q_- F_4(w) &= -\frac{\lambda_B}{(2\pi)^2}\int \frac{d^2 q'}{(2\pi)^2} \chi(w')\frac{(1 - F_4)^2 q^{\prime 2}_-}{(q+q')_-}\ .
\end{align}
In the next section we will frequently require the (indefinite)
integral of $\chi(z)$ defined in \eqref{intdet} with respect to
$z$. This integral is easily evaluated; we find
\begin{align}\label{intchi}
   \xi(z) &= 
   -\frac{1}{2(2\pi)^3}\int^z dw' \chi(w')\ ,\\
   &= \frac{1}{2 m \beta }\int d\alpha \rho_B(\alpha) \bigg[\log 2\sinh(\tfrac{\beta}{2}(\sqrt{z+M^2(z)}+i\tfrac{\alpha}{\beta})) +\nonumber \\ &\qquad\qquad\qquad\qquad\qquad\qquad + \log 2\sinh(\tfrac{\beta}{2}(\sqrt{z+M^2(z)}-i\tfrac{\alpha}{\beta})) \bigg]\nonumber\ .
\end{align}

\subsection{Dimensional regularization}\label{dimreg}
Later in this section we will encounter divergent integrals that 
will need to be regulated. Following \cite{Giombi:2011kc} we will perform this regulation by employing dimensionally regulated version 
of  the summation in \eqref{intdet} which effectively 
replaces \eqref{intdet} by  (see around Sec (2.2) of \cite{Giombi:2011kc} for details)
\begin{multline}\label{regchi}
\chi(z)= -\frac{2\pi^3}{m} \int d\alpha \rho_B(\alpha) \frac{1}{(\sqrt{z + M^2(z)})^{1+\epsilon}}\times \\ \times\left[\coth (\tfrac{\beta}{2}(\sqrt{z + M^2(z)} + i\tfrac{\alpha}{\beta})) + \coth (\tfrac{\beta}{2}(\sqrt{z + M^2(z)} - i\tfrac{\alpha}{\beta}))\right]\ ,
\end{multline}
and effectively replaces \eqref{intchi} by
\begin{multline}\label{regintchi}
  \xi(z) = \frac{1}{2 m \beta}\int d\alpha \rho_B(\alpha) \frac{1}{(\sqrt{z+M^2(z)})^\epsilon}\times \\ \times\left[\log 2\sinh(\tfrac{\beta}{2}(\sqrt{z+M^2(z)}+i\tfrac{\alpha}{\beta}))+\log 2\sinh(\tfrac{\beta}{2}(\sqrt{z+M^2(z)}-i\tfrac{\alpha}{\beta})) \right]\ ,
\end{multline}
where $\epsilon$ is an infinitesimal that is taken to zero at the end
of the computation (there are additional terms in the indefinite
integral of \eqref{regchi} which are proportional to
$\epsilon$. However, since we always take the limit $\epsilon \to 0$,
we drop these extra terms from \eqref{regintchi}).

\subsection{Performing the angular integrals}
While the RHS of \eqref{inteqexp2} is given in terms of two
dimensional integrals $d^2 q'$ the unknown functions on the RHS are
functions only of $w'$. For this reason we can explicitly perform all
angular integrals in the RHS of the four equations \eqref{inteqexp2}.
Let $q'_- = \tfrac{1}{\sqrt{2}} q'_s e^{i\eta}$.  It is easy to check
that
\begin{align}
&\int_0^{2\pi} \frac{d\eta}{2\pi} \frac{1}{q_- + q'_-} = \frac{1}{q_-} \Theta(q_s - q'_s)\ ,\quad \int_0^{2\pi} \frac{d\eta}{2\pi} \frac{1}{(q_- + q'_-)q'_-} = -\frac{1}{q^2_-}\Theta(q_s - q'_s)\ ,\\
&\int_0^{2\pi} \frac{d\eta}{2\pi} \frac{q'_-}{q_- + q'_-} = \Theta(q'_s - q_s)\ ,\quad \int_0^{2\pi} \frac{d\eta}{2\pi} \frac{q^{\prime 2}_-}{q_- + q'_-} = -q_-\Theta(q'_s - q_s)\ .
\end{align}
Using these results to perform all angular integrals on the RHS of \eqref{inteqexp2} we find that the integral equations 
take the form
\begin{align} \label{inteqexp3}
 F_1(w) &= -\frac{\lambda_B}{(2\pi)^2}\int_0^{w} \frac{dw'}{4\pi} \chi(w')\left(2F_1 (1-F_4) + (F_2 + im)(2 F_3 + w')\right)\ ,\nonumber\\
  F_2(w) &= \frac{\lambda_B}{(2\pi)^2}\int_w^\infty \frac{d w'}{4\pi} \chi(w') (1 - F_4) (F_2 + im)\ ,\nonumber\\
  F_3(w) &= \frac{\lambda_B}{(2\pi)^2} \int_0^w \frac{d w'}{4\pi} \chi(w')\left((F_3 + \tfrac{w'}{2})(1 - F_4) - im( F_2 + im)\right)\ ,\nonumber\\
  F_4(w) &= \frac{\lambda_B}{(2\pi)^2}\int_w^\infty \frac{d w'}{4\pi} \chi(w') (1 - F_4)^2\ .
\end{align}

\subsection{Differential equations for the unknowns}
We obtain the following differential equations for the functions
$F_1(w), \ldots, F_4(w)$ by differentiating them with respect to their
arguments:
\begin{equation}
\label{dieq}\begin{split} 
F_1'(w)& = -\frac{\lambda_B}{2(2\pi)^3} \chi(w) \left(2F_1 (1-F_4) + (F_2 + im)(2 F_3 + w)\right)\ ,\\
F'_2(w)& = -\frac{\lambda_B}{2(2\pi)^3} \chi(w) (1-F_4)(F_2+im)\ , \\
F'_3(w)& = \frac{\lambda_B}{2(2\pi)^3} \chi(w) \left((F_3 + \tfrac{w}{2})(1 - F_4) - i m(F_2 + im)\right)\ ,\\
F'_4(w)& = -\frac{\lambda_B}{2(2\pi)^3} \chi(w)  (1 - F_4)^2\ .
\end{split}
\end{equation}
Now it follows from the definition of $M^2(w)$ in \eqref{detQ} that when (\ref{dieq}) are satisfied
\begin{equation} \label{mdeif}
 M'(w) =-\frac{1}{2 M(w)} \left(F_4(w)+\frac{i F_2(w) (1-F_4(w))}{m}\right)\ .
\end{equation} 
We will return to this equation in a bit. 

\subsection{Solving the gap equations} 
\subsubsection{Determining \texorpdfstring{$F_4$}{F4}} \label{dff}
The integral equation for the function $F_4$ (see \eqref{inteqexp3})
can be solved very simply in multiple different ways. To begin with we
solve this equation order by order in perturbation theory. We proceed
by expanding $F_4$ in a perturbative expansion in $\lambda_B$
\begin{equation}
  F_4 = f_{0} + \lambda_B f_{1} + \lambda_B^2 f_{2} + \cdots\ .
\end{equation}
and simply plug this ansatz back in to the integral equation. 
The equation takes the form 
\begin{equation}
  f_0 + \lambda_B f_1 + \lambda_B^2 f_2 + \cdots = \frac{\lambda_B}{2(2\pi)^3}\int_w^\infty dw' \chi(w') (f_0 - 1 + \lambda_B f_1 + \lambda_B^2 f_2 + \cdots)^2\ .
\end{equation}
and yields the following infinite sequence of equations 
\begin{equation} \label{secsol} 
  f_0 = 0\ ,\quad f_1 = \frac{1}{2(2\pi)^3}\int_w^\infty dw' \chi(w')\ ,\quad f_2 = -\frac{1}{2(2\pi)^3}\int_w^\infty dw' \chi(w') 2f_1(w')\ ,\ldots
\end{equation}
Each of these equations can be solved in a straightforward manner by
integration. The only subtlety here is that, at every order, the
indefinite integrals in question diverge at large $w'$. We use the
dimensional regulation scheme outlined in subsection \ref{dimreg} to
define these divergent integrals.  In order to proceed with our
analysis we assume that the mass parameter $M(z)$ tends to a constant
$M$ at least at large $z$; in the next subsection we will demonstrate
that this assumption is self-consistent.

In order to see how this works lets start with the 
second equation in \eqref{secsol}. It follows 
immediately from the definition \eqref{intchi} that 
\begin{equation}
\label{fone} 
f_1= {\xi}(w) - {\xi }(\infty)\ .
\end{equation}
The problem with \eqref{fone} is that ${\xi}(\infty)$ is divergent;
indeed it is easily verified from the regularized version
\eqref{regintchi} that at large $w$
\begin{equation}
  \xi(w) \to \frac{1}{2m} (\sqrt{w + M^2(w)})^{1-\epsilon} 
  \equiv {\xi}_{\text{asymp}}(w)\ .
\end{equation}
In order to make sense of \eqref{fone} we proceed as follows. 
Consider the function $B(w)$ 
$$B(w) = \frac{1-\epsilon}{4 m (\sqrt{w + M^2(w)})^{1+\epsilon} }
$$
$B(z)$ is defined so as to obey the identity 
$$ B(z)= {\xi}_{\text{asymp}}'(z)$$
Now we evaluate the integral for $f_1(w)$ in \eqref{secsol} as follows:
\begin{equation} \label{secosn}
-2(2\pi)^3f_1 = \int_w^\infty dw'  \left( \chi(w') - B(w') \right) + 
\int_w^\infty B(w') 
\end{equation}
The first integral in \eqref{secosn} is now convergent and 
evaluates to 
$$ -{\xi}(w) + {\xi}_{\text{asymp}}(w)\ ,$$
(note that the contributions at infinity cancel). The second integral
in \eqref{secosn} is divergent and is evaluated using dimensional
regularization. It evaluates to
$$ -{\xi}_{\text{asymp}}(w)\ .$$
Adding together the two terms we find the well defined expression
\begin{equation}
\label{fonem} 
f_1(w)= {\xi}(w)\ .
\end{equation}
Notice that the net effect of our dimensional regulation scheme was to
simply drop the surface term at infinity. It is easy to convince
oneself that this scheme effectively does the same thing (i.e. drops
all surface terms at infinity) in all the integrals that appear in the
perturbative evaluation of $F_4$.  Adopting this prescription we find
%Thus, the integrals for $f_1$, $f_2$ etc.~are divergent. We adopt a
%regularization scheme where we drop terms proportional to powers of
%$\sqrt{z_\infty + M^2}$. Then, we get
%\begin{equation}
%  f_0 = 0\ ,\quad f_1 = - \xi(w)\ ,\quad f_2 = -\xi(w)^2\ ,\ \ldots %\frac{2}{(2\pi)^3}\int_w^\infty dw' \xi'(w') \xi(w') = \ ,\ldots
%\end{equation}
%
% The equation for $f_3$ is
% \begin{equation}
%   f_3 = -\frac{1}{2(2\pi)^3} \int_w^\infty dw' \chi(w') (-2 f_2 + f_1^2) = -\int_w^\infty dw' \xi'(w') 3\xi(w')^2 = \xi(w)^3\ .
% \end{equation}
%
% It is straightforward to see that the perturbative series is given by
\begin{align} \label{sumf4}
  F_4(w) &=\lambda_B \xi(w) - \lambda_B^2 \xi(w)^2 + \lambda_B^3 \xi(w)^3 - \cdots\nonumber\\
      &= 1 - \sum_{n=0}^\infty (-\lambda_B)^n\, \xi(w)^n = 1 - \frac{1}{1 + \lambda_B \xi(w)}\ .
\end{align}
Summing up, we have
\begin{equation}\label{F4pert}
  1- F_4 = \frac{1}{1 + \lambda_B \xi(w)}\ .
\end{equation}
As a consistency check, it is easy to verify that our solution 
\eqref{F4pert} obeys its differential equation 
(fourth of \eqref{dieq}). Indeed this differential equation is 
easy to solve in generality; its most general solution is 
\begin{equation}\label{F4sol}
\frac{1}{1 - F_4(w)} = \lambda_B \xi(w) + c_4\ ,
\end{equation}
where $c_4$ is an integration constant. Clearly \eqref{F4sol} 
reduces to \eqref{F4pert} if we choose
\begin{equation}\label{c4sol}
  c_4 = 1\ .
\end{equation}

\subsubsection{A subtlety in \texorpdfstring{$F_4$}{F4}}

In order to complete the process of checking our solution, 
let us directly check that the solution \eqref{F4pert} obeys 
the integral equation \eqref{inteqexp3} which we 
reproduce here for clarity. 
\begin{equation}
\label{ffeq}
F_4(w) = \frac{\lambda_B}{(2\pi)^2}\int_w^\infty \frac{d w'}{4\pi} \chi(w') (1 - F_4)^2
\end{equation}
We will find that this check helps us better understand the procedure we used to obtain the  solution \eqref{F4pert}, by contrasting 
it with an equally reasonable sounding procedure that does not work.

Let $F_4$ be any solution of the fourth of \eqref{dieq}, i.e. 
a solution of the form \eqref{F4sol} with any value of $c_4$.
For every such solution  
\begin{equation}
\label{f4t}
(1-F_4)^2= \frac{1}{(\lambda_B \xi(w) + c_4)^2}
\end{equation}
Inserting this into the RHS of \eqref{ffeq} and using the fact that 
\begin{equation} \label{lkl}
\chi(w) dw = -2(2 \pi)^3 d \xi
\end{equation} 
we conclude  that the RHS of \eqref{ffeq} evaluates to 
$$- \frac{1}{\lambda_B \xi(w) + c_4}= F_4-1$$
(where we have used the fact that $\xi$ diverges as $w \to \infty$
- note in particular that the integral on the RHS of \eqref{ffeq} 
is convergent).
On the other hand the LHS of \eqref{ffeq} is $F_4$. As 
$F_4 \neq F_4-1$ we find that the RHS and LHS of these equations 
do not agree for any value of $c_4$. It follows, in other words, 
that {\it no} solution of the differential equations \eqref{dieq} 
obeys the integral equation \eqref{ffeq}. As we have 
earlier argued that every solution of the integral equations 
(e.g. \eqref{ffeq}) obey the differential equations \eqref{dieq}, 
we are forced to conclude that the integral equation 
\eqref{ffeq} has no solutions! 

The conclusion of the previous paragraph appears to be in 
direct conflict with the fact that - in the previous subsubsection 
- we have actually found an explicit solution - 
namely \eqref{F4pert} - of the equation \eqref{ffeq}. To make 
this contradiction as sharp as possible let us specialize the 
analysis in the paragraph around \eqref{f4t} to the special 
case $c_4=1$. In this case the solution presented in \eqref{f4t}
is the perturbative solution \eqref{F4pert}. How can it be that 
the analysis in the paragraph around \eqref{f4t} demonstrates 
that this solution does not obey the integral equation 
\eqref{ffeq}, while the analysis earlier in this subsection 
demonstrates that it does?

The answer to this question is simply that the expansion of 
the quantity $(1-F_4)^2$ on the RHS of \eqref{ffeq} in a power 
series in $\lambda_B$ does not commute with the integral over 
$w'$ in \eqref{ffeq}. More precisely let us contrast two methods of evaluating the integral \eqref{ffeq} that give different 
answers. 

The first method - the one adopted in this subsubsection - is to
performing the sum over $\lambda_B$ first (as in \eqref{sumf4}), then
to notice that the resultant integrand defines a convergent integral
in \eqref{ffeq}, and to evaluate the integral.

The second method - adopted in subsubsection \ref{dff} - on the other
hand, is to first expand the integrand on the RHS of \eqref{ffeq} in a
power series in $\lambda_B$, perform the integral order by order for
each of the coefficients of $\lambda_B^n$ and then to sum the final
power series of results. Crucially the integrals encountered at every
order in the $\lambda_B$ expansion are divergent and need to be
defined.  Defining the integrals by a form of dimensional
regularization yields a result for the RHS that agrees with the LHS of
\eqref{ffeq}.

The second method is guaranteed to reproduce the results of Feynman
diagram based perturbation theory (because it simply is the integral
equation's way of generating Feynman diagrams loop by loop). As we
require all our results to agree with perturbation theory we will take
the view that the second method is the correct one all through this
paper, and so \eqref{F4pert} is the correct solution for $F_4$.

We will not encounter similar subtleties in any of the other integral equations in this paper. 

\subsubsection{A curious observation}

In this subsection we note a curious fact relating to the subtlety
of the last subsection. It turns out that there is a second, apparently ad hoc - but nonetheless interesting procedure that yields the same answer for our $W$ boson propagator as the procedure outlined 
in the previous subsection and employed in the rest of this paper. This subsection is devoted to a 
description of this alternate procedure. 

The analysis of this subsection will be used no where else in this 
paper - and may turn out to be a curiosity with no 
deeper significance. The impatient reader should feel free to 
skip over to the next subsection.

The ad hoc procedure we will employ in this subsection is to modify our starting action - \eqref{lcg} - in the 
manner that we now describe: we simply drop the term proportional 
to $W_+ \p_- W_3$ that occurs in the expansion of the first term in the 
second line of \eqref{lcg}. If we then  rerun the analysis of this 
paper but starting with this modified action we find, in particular, 
that 
\begin{equation}   \label{qitfoffn}
Q^{\mu\nu}(q)=\frac{1}{2\pi}\begin{bmatrix}
0  & -i( F_2 + im + q_3)   & i q_-( - {\tilde F}_{4})  \\
-i( F_{2} + im - q_3) &  -\frac{i}{q_-^2} F_{1}(w) & -\frac{i}{q_-}(F_3 + \tfrac{w}{2})\\
-i q_-(- {\tilde F}_{4}) & \frac{i}{q_-}(F_{3} + \tfrac{w}{2}) &  m
\end{bmatrix}\ .
\end{equation}
where ${\tilde F}_4$ parametrizes the self energy contribution 
to $\Sigma^{3 +}$ in the modified problem, in exactly the same 
way that $F_4$ parametrized the same quantity in the original 
problem. \footnote{In principle we should also replace $F_2$ by 
	${\tilde F}_2$, and similarly for $F_3$ and $F_1$, but it 
	will turn out below that ${\tilde F}_i=F_i$ for $i= 1 \ldots 3$, so 	we will avoid cluttering the notation. }
Proceeding as above, we find that our modified problem leads 
to the integral equations 
\begin{align} \label{inteqexp3n}
F_1(w) &= -\frac{\lambda_B}{(2\pi)^2}\int_0^{w} \frac{dw'}{4\pi} \chi(w')\left( 2F_1(- {\tilde F}_4)  + (F_2 + im)(2 F_3 + w')\right)\ ,\nonumber\\
F_2(w) &= \frac{\lambda_B}{(2\pi)^2}\int_w^\infty \frac{d w'}{4\pi} \chi(w') (- \tilde F_4) (F_2 + im)\ ,\nonumber\\
F_3(w) &= \frac{\lambda_B}{(2\pi)^2} \int_0^w \frac{d w'}{4\pi} \chi(w')\left((F_3 + \tfrac{w'}{2})(-\tilde F_4) - im( F_2 + im)\right)\ ,\nonumber\\
\tilde F_4(w) &= \frac{\lambda_B}{(2\pi)^2}\int_w^\infty \frac{d w'}{4\pi} \chi(w') (- \tilde F_4)^2\ .
\end{align}
 
In particular the last of \eqref{inteqexp3n} is easily solved 
at finite $\lambda_B$ (there is not need to expand in $\lambda_B$ 
before performing the integrals) and we find 
\begin{equation}
\label{tff}
{\tilde F}_4=F_4-1= - \frac{1}{1+ \lambda_B \xi(w)}
\end{equation}
Comparing \eqref{qitfoffn}, \eqref{qitfoff} and \eqref{tff}, 
it follows that $Q^{\mu\nu}$ of this subsection is now identical 
- as a function of $F_2$, $F_3$ and $F_1$ - to $Q^{\mu\nu}$ 
of the actual problem. The remaining integral equations of modified 
problem -  the first three of \eqref{inteqexp3n} with 
\eqref{tff} plugged in - are now identical to the integral equations 
\eqref{inteqexp3} of the original problem with \eqref{F4pert} plugged 
in. It follows, in particular, that from this point on, the 
equations for the two problems are the same.

Let us summarize. There are two procedures that yield the same 
thermal propagator. The first uses the actual classical action 
of our system as its starting point but evaluates all integrals 
by expanding out the integrands term by term in an expansion in $\lambda_B$ and then evaluating the integrals that appear 
at each order using the dimensional regularization scheme. This is the procedure adopted in earlier subsections and 
in the rest of this paper. In the second procedure we evaluate
all integrals first (before performing any expansions in $\lambda_B$ 
that may be of interest). The second procedure gives us the same 
result as the first, if we modify the starting action with a very particular `counterterm'.

In other words it appears that the two different regulation schemes 
(using dimensional regularization before or after expanding in $\lambda_B$)
differ by a very particular counterterm. It is, of course, usual for different regulation schemes to effectively
differ by counterterms. The novelty in the current situation is that 
the needed counterterm occurs at leading (classical) order in the 
loop expansion rather than at higher orders as is more usual.

\subsubsection{Determining \texorpdfstring{$F_2$}{F2}}
We will now determine the function $F_2$. The differential equation
for $F_2$ is given by
\begin{equation}
  F'_2(w) = -\frac{\lambda_B}{2(2\pi)^3} \chi(w) (1-F_4)(F_2+im)\ .
\end{equation}
Plugging in the expression for $1-F_4$ from \eqref{F4pert} and using \eqref{intchi}, we have
\begin{equation}
  \frac{d(F_2+im)}{F_2+im} = \frac{d(1 + \lambda_B \xi(w))}{1 + \lambda_B \xi(w)}\ ,
\end{equation}
which gives the solution
\begin{equation}\label{F2desol}
  F_2(w) + im = c_2 (1 + \lambda_B \xi(w))\ ,
\end{equation}
where $c_2$ is an integration constant.

In order to determine the constant $c_2$ we plug \eqref{F2desol} 
into the second of \eqref{inteqexp3}. Using \eqref{lkl} we 
see that the RHS of that integral equation evaluates to
\begin{equation}
\label{fweq} 
- c_2 \lambda_B \int_w^\infty   d \xi(w)
\end{equation}
The integral \eqref{fweq} is divergent and must be evaluated 
after dimensional regularization. Exactly as in subsubsection 
\ref{dff}, the net result of this regulation scheme is to 
simply drop the surface term at infinity. We conclude that the 
integral in \eqref{fweq} evaluates to 
$$ c_2 \lambda_B \ \xi(w).$$ 
The integral equation is satisfied if this expression also equals 
$F_2$. Using \eqref{F2desol} this condition takes the form
\begin{equation}\label{c2det}
c_2 (1 + \lambda_B \xi(w)) -im = c_2 \lambda_B \xi(w)\quad {\rm i.e.}\quad c_2= im\ .
\end{equation}
We conclude that the unknown function $F_2$ is given by 
\begin{equation}
\label{finalf2}
F_2(w)= i m \lambda_B \xi(w)\ .
\end{equation}

\subsubsection{Determination of the mass} 
Plugging \eqref{finalf2} and \eqref{F4pert} into \eqref{mdeif} 
we find 
\begin{equation} \label{mevol}
M'(w)=0,
\end{equation}
In other words the complicated mass function $M(w)$, listed in 
\eqref{detQ} is just a constant independent of $w$. We pause 
to recall why this result is extremely satisfying, both from 
the physical and the technical point of view. 

Recall  that poles of the $W$ boson lie at the zeroes of the determinant of $Q$. 
Now the poles of $W$ boson particles have gauge invariant physical content 
(they determine the dispersion relation of the $W$ bosons). At zero temperature we  expect 
this dispersion to be Lorentz invariant. It follows from \eqref{detQ}  that 
this is only the case if $M$ is a constant independent 
of $w$. The fact that $M$ comes out to be constant and serves as a nontrivial consistency check on our 
results at zero temperature\footnote{Note that the full $W$ boson propagator - which 
is gauge dependent and so unphysical - is far from Lorentz invariant in our gauge. 
It is gratifying that, nonetheless, the gauge invariant data in the propagator  
is Lorentz invariant.}.

It is not immediately clear that there is a clear physical reason to
expect that $M$ had to be constant, independent of $w$, even away from
the zero temperature limit. However the fact that this turns out to be
the case is satisfying for two reasons.  First, it allows us to give a
clear interpretation to the quantity $M$; $M$ is the `thermal mass' of
the $W$ bosons. More importantly, at the technical level, the fact
that $M$ is a constant turns the function $\chi$ into a completely
known function of $w$ (it was previously known in terms of the unknown
function $M(w)$).  This fact turns the differential equations for
$F_3$ and $F_1$ into linear differential equations that are easily
solved. We will return to this point in the next subsubsection.

Of course the constant value of the mass $M$ is not a free parameter;
it is itself determined in terms of the parameters of the theory and
the temperature. In the rest of this subsubsection we will find an
equation that determines the value of $M$.

Inserting the relation
\begin{equation}
  -\frac{i}{m} (F_2 + im) (1-F_4) = 1\ ,
\end{equation}
into the expression for $M^2$ in \eqref{detQ} we conclude that 
\begin{equation}\label{SO3mass}
  M^2 = -(F_2 + im)^2 - \tfrac{i}{m}F_1 (1-F_4)^2 + 2F_3\ .
\end{equation}
\eqref{SO3mass} is a functional relationship that holds at every 
value of $w$. The RHS of \eqref{SO3mass} involves the functions 
$F_3$ and $F_1$ that we still do not know at general values 
of $w$. However the structure of the last two equations 
\eqref{inteqexp2} ensures that $F_3(0)=F_1(0)=0$. Evaluating 
\eqref{SO3mass} at $w=0$ we find the equation 
\begin{equation}\label{bge}
  M^2 = -(F_2(0) + im)^2 = m^2 \left(\lambda_B \xi(0) + 1\right)^2\ .
\end{equation}
Recall
\begin{equation}\label{xi0}
 \xi(0) = \frac{1}{2\beta m }\int d\alpha \rho_B(\alpha) \left[\log 2\sinh(\tfrac{\beta}{2}(M+i\tfrac{\alpha}{\beta}))+\log 2\sinh(\tfrac{\beta}{2}(M-i\tfrac{\alpha}{\beta})) \right]= \frac{\mathcal{S}}{m \beta} \ .
\end{equation}
where, we recall from \eqref{ss}, that $\mc{S} = \mc{S}(M, 0)$. To
proceed, we define the dimensionless quantities
\begin{equation}\label{cBdef}
c_B = \beta M\ ,\quad \hat{m} = \beta m\ .
\end{equation}
It follows that \eqref{bge} can be recast in terms of $c_B$ and
$\hat{m}$ into the equation
\begin{equation}
  c_B^2 = \left(\hat{m} + \frac{\lambda_B}{2}\int_{-\pi}^\pi d\alpha \rho(\alpha) \left[\log 2\sinh\left(\tfrac{c_B+i\alpha}{2}\right)+\log 2\sinh\left(\tfrac{c_B-i\alpha}{2}\right) \right]\right)^2\ .
\end{equation}
Using \eqref{xi0} and the expression for $\hat{m}$ in terms of
$\hat{m}_B^{\text{cri}}$ from \eqref{convn}, the above may be rewritten as
\begin{equation} \label{fge} 
  (2c_B)^2 = \left(-\lambda_B \hat m_B^{\text{cri}} + 2\lambda_B \cS \right)^2=\left(-|\lambda_B| \hat m_B^{\text{cri}} + 2|\lambda_B| \cS \right)^2
\end{equation}
This is our final gap equation for the thermal mass of the $W$ bosons.

\subsubsection{Solving for \texorpdfstring{$F_3$}{F3} and \texorpdfstring{$F_1$}{F1}}
We next focus on the differential equations for $F_1$ and $F_3$. We
define the function $g(w)$:
\begin{equation}\label{defg}
g(w) = \lambda_B \xi(w) + 1 =  \frac{1}{im}(F_2 + im) = \frac{1}{1 - F_4}\ ,\quad\text{with}\quad g'(w) = -\frac{\lambda_B}{2(2\pi)^3} \chi(w)\ .
\end{equation}
Then, the differential equation for $F_3$ becomes
\begin{equation}\label{def3}
F'_3(w) + \frac{g'(w)}{g(w)} (F_3(w) + \tfrac{w}{2}) + m^2 g'(w)g(w) = 0\ ,
\end{equation}
which we rewrite as
\begin{equation} \label{ftode}
\begin{aligned} 
 (F'_3 & (w) + \tfrac{1}{2}) + \frac{g'(w)}{g(w)} (F_3(w) + \tfrac{w}{2}) + m^2 g'(w)g(w) - \tfrac{1}{2} = 0\\
& ((F_3(w) + \tfrac{w}{2})g(w))' + m^2 g'(w)g(w)^2 - \tfrac{1}{2}g(w) = 0
\end{aligned}
\end{equation}
Integral equation for $F_3$ requires $F_3(0)=0$. With this boundary condition above equation can be integrated to give
\begin{equation}
	\begin{aligned}
		F_3(w) =- \frac{w}{2} +\frac{1}{g(w)}\left(\frac{1}{2}\mathcal{I}(w)-\frac{m^2}{3} ( g(w)^3-g(0)^3   ) \right)
	\end{aligned}
\end{equation}
where we have defined
\begin{equation} \label{defI}
	\begin{aligned}
		\mathcal{I}(w)=\int_{0}^{w}g(z)dz
	\end{aligned}
\end{equation}

To get a simplified equation for $F_1$ we use (\ref{detQ}) to eliminate $F_3$ from RHS of (\ref{dieq}  ) ($M^2$ is independent of $w$ as follows from previous discussions) to give a differential equation for $F_1$
\begin{equation}
\begin{aligned}
	 F_1'(w)& - \frac{g'(w)}{g(w)} F_1(w) + im g'(w) g(w) (g(w)^2 m^2-M^2-w) = 0\\ 
& \left( \frac{F_1(w)}{g(w)}\right)' + im g'(w)(g(w)^2 m^2-M^2-w) = 0
\end{aligned}
\end{equation}
Integral equation for $F_1$ requires $F_1(0)=0$. With this boundary condition above equation can be integrated to give
\begin{equation}
	\begin{aligned}
		F_1(w) & = im g(w) \left( M^2 (g(w)-g(0))-\frac{m^2}{3}(g(w)^3-g(0)^3)+\int_0^w z g'(z) dz \right)\\
		&= im g(w) \left( M^2 (g(w)-g(0))-\frac{m^2}{3}(g(w)^3-g(0)^3)+ w g(w)-\mathcal{I}(w)  \right)
	\end{aligned}
\end{equation}
Using formula for mass \ref{bge}, it can be easily checked that these
solutions indeed satisfy \ref{detQ} as required by consistency. We
next plug these solutions into the effective action and obtain the
free energy functional $v_B$. Before that, we take a short digression
and discuss the case with chemical potential.

\subsubsection{Adding a Chemical Potential}
The $SU(N_B)$ theory \eqref{basiclag} enjoys invariance
under a global $U(1)$ symmetry. The action of the $U(1)$ global
symmetry element $e^{i \alpha}$ on the fundamental multiplet is given
by
\begin{equation}\label{uoph}
\phi \rightarrow e^{i \alpha} U(\alpha) \phi, ~~~{\bar \phi } \rightarrow {\bar \phi} U^\dagger(\alpha) e^{-i \alpha}
, ~~~A_\mu \rightarrow U(\alpha) A_\mu U^\dagger(\alpha)
\end{equation}
where $U(\alpha)$ is any one parameter choice of  $SU(N_B)$ gauge transformations. As $U(\alpha)$ generate gauge 
transformations, different choices of $U(\alpha)$ all actually generate the same symmetry. The matrix $U(\alpha)$
can be chosen in any convenient manner. 

We will find it convenient to choose $U(\alpha)$ to be given by 
\begin{equation} \label{ualpha}
U(\alpha)= {\rm Diag} \left( e^{i \frac{\alpha}{N_B-1}}, e^{i \frac{\alpha}{N_B-1}}, \ldots, e^{i \frac{\alpha}{N_B-1}}, 
e^{-i \alpha}  \right)
\end{equation} 
With this choice the action \eqref{uoph} preserves our unitary gauge choice \eqref{sqv}. As we have explained above, this 
choice of gauge freezes out all $\phi$ degrees of freedom. The gauge fields of the 
unbroken $SU(N_B-1)$ gauge group transform trivially under \eqref{uoph}, as do the neutral $Z$ bosons. On the other 
hand the fundamental $W$ bosons transform under \eqref{ualpha} like objects of charge $-\frac{N_B}{N_B-1}$. In the 
large $N_B$ limit under study in this paper, the charge of these $W$ bosons is just $-1$. 

The reader may find herself surprised by the fact that the global charge of a $W_\mu$ boson differs (although ever so slightly) from $-1$. In fact this 
charge renormalization is actually natural in the 
$SU(N_B)$ theory with which we are working. Recall 
that neither $\phi$ (in the unHiggsed phase) nor 
$W_\mu$ (in the Higgsed phase) are gauge invariant 
operators. We can build gauge invariant operators 
by contracting $\phi$ (or $W$) with their complex conjugates, but such operators carry no global charge. The simplest charged gauge invariant operator is a `baryon' build by contracting 
$N_B$ $\phi $ \footnote{In our schematic discussion we use the same notation for $\phi$ and any of its derivatives; similarly for $W_\mu$. } 
or $N_B -1$ $W_\mu$ operators. In the unHiggsed 
phase such a baryon clearly carries global symmetry charge 
$N_B$. This precisely matches the global symmetry charge of a baryon operator made out of $N_B-1$ 
${\bar W}_\mu$ fields precisely {\it because of}
the charge `renormalization' described above. 
In other words the charge renormalization is precisely what is needed in order to ensure that the charges of gauge invariant operators do not jump as we 
move from the Higgsed to the unHiggsed phase.	
\footnote{Note that the phenomenon we have just explained - namely the `renormalization' of 
global charge - does not occur in the $U(N_B)$ theory in which case $W_\mu$ bosons have global charge $-1$. This matches with the fact that the explanation we have presented also does not 
apply to the $U(N_B)$ theory which has no 
baryonic operators. }

There is another way of understanding fact that the ratio of the magnitude 
	of the charge of a $W$ boson and the original $\phi$ field 
	is $\frac{N_B}{N_B-1}$ using duality. Let $B_\mu$ be the background gauge field that 
	couples to the global $U(1)$ symmetry of the $U(N_F)_{k_F}$ 
	fermionic theory. The coupling in question is proportional 
	to $\int B d a$ where $a$ is the $U(1)$ part of the dynamical
	$U(N_F)$ fermionic gauge field. A single fermionic particle traps $da$ flux
    proportional to $\frac{1}{|k_F|}$ when $m_F$ and $k_F$ have
    the same sign (i.e. in the dual of the un Higgsed phase) 
    but $da$ flux proportional to $\frac{1}{|k_F|-1}$ when 
    $m_F$ and $k_F$ have opposite signs, i.e. in the dual 
    to of the Higgsed phase. It follows that the ratio of charges
    of excitations in the (fermionic dual to) the unHiggsed 
    and Higgsed phases is $\frac{|k_F|}{|k_F|-1}$, which 
    exactly maps to  $\frac{N_B}{N_B-1}$ under duality.

Let us now repeat the computation of the thermal partition function, presented earlier in this section, after turning 
on a chemical potential $\mu$ for the $U(1)$ charge with action listed in \eqref{uoph}. This may be accomplished 
by turning on an imaginary background gauge field with $ A_0 =i \mu$ for all fundamental $W$ fields and
an imaginary gauge field with $ A_0=-i\mu$ for the antifundamental ${\bar W}$ fields. This is achieved by 
making the replacement $\alpha \rightarrow \alpha -i  \nu $ in \eqref{intdet} and \eqref{intchi} etc. 
In other words we define generalized $\chi$ and $\xi$ functions by the formulae
\begin{align}\label{intdetg}
  \chi(z) &\equiv -\frac{8\pi^3}{m} \int d\alpha \rho_B(\alpha)\, \beta^{-1}\sum_{n\,\in\,\mathbb{Z}} \frac{1}{(2 \pi \frac{n}{\beta}+\frac{\alpha}{\beta} -i\frac{\nu}{\beta})^2+(z+M^2(z))}\ ,\nonumber\\
    %      &= \frac{2\pi\beta}{m} \int d\alpha \rho_B(\alpha)\sum_{n\,\in\,\mathbb{Z}} \frac{1}{( n + \tfrac{\beta}{2\pi}\frac{\alpha}{\beta})^2+ \tfrac{\beta^2}{4\pi^2}(z+M^2(z))}\ ,\nonumber\\
          &= -\frac{2\pi^3}{m} \int d\alpha \rho_B(\alpha) \frac{1}{\sqrt{z + M^2}}\bigg(\coth (\tfrac{\beta}{2}(\sqrt{z + M^2} + i\tfrac{\alpha}{\beta} +\tfrac{\nu}{\beta})) \nonumber\\ & \qquad\qquad\qquad\qquad\qquad\qquad\qquad+ \coth (\tfrac{\beta}{2}(\sqrt{z + M^2} - i\tfrac{\alpha}{\beta} -\tfrac{\nu}{\beta}))\bigg)
\end{align}
and
\begin{align}\label{intchig}
   \xi(z) &= -
   \frac{1}{2(2\pi)^3}\int^z dw' \chi(w')\ ,\\
   &= \frac{1}{2\beta m}\int d\alpha \rho_B(\alpha) \bigg(\log 2\sinh(\tfrac{\beta}{2}(\sqrt{z+M^2}+i\tfrac{\alpha}{\beta} + \tfrac{\nu}{\beta}))\nonumber \\ & \qquad\qquad\qquad\qquad\qquad\qquad\qquad +\log 2\sinh(\tfrac{\beta}{2}(\sqrt{z+M^2}-i\tfrac{\alpha}{\beta} -\tfrac{\nu}{\beta})) \bigg)\nonumber\ .
\end{align}

Earlier in this subsection we have obtained explicit results for the thermal self energy and propagators of our theory. 
Our results were expressed in terms of the functions $F_1 \ldots F_4$ which, in turn, we solved for in term of 
$\chi$ and $\xi$. All of these results also go through in the presence of a chemical potential if we replace 
the functions $\chi$ and $\xi$ with their generalizations defined in this subsection.

\section{The Free energy}
Recall that the free energy functional $v_B$ is given in the saddle
point approximation by the effective action for the $\alpha$ and
$\Sigma$ fields (cf.~\eqref{naab}):
\begin{equation} \label{Seff}
\begin{split}
\mc{V}_2 T^2 v_B(|c_B|,\rho_B) = S_{\text{eff}}[\alpha,\Sigma] &= {N_B \mathcal{V}_3}\bigg(V_0[\alpha] -\frac{i}{\lambda_B} \ \int 
\frac{\td^3q}{(2 \pi)^3}
\Sigma^{\nu\mu}(q) \alpha_{\mu\nu}(-q) \\ & \qquad\qquad\qquad + \int \frac{\td^3q}{(2 \pi)^3} \log {\rm det}
\left(K_W(q)  -i \Sigma^{T}(q) \right) \bigg)\ .
\end{split}
\end{equation}
The quantity $V_0[\alpha]$ in \eqref{Seff} is complicated because it 
involves a double integral over momenta even when evaluated 
on translationally invariant solutions. On shell, however, it is 
possible to eliminate $V_0[\alpha]$ using the equations of motion. 
Using the fact that $V_0[\alpha]$ is a homogeneous polynomial of degree 2 in $\alpha$ it follows (from the $\alpha$ equation of motion)
  that, on-shell, 
  \begin{equation}
  	\begin{aligned}
  		V_0[\alpha]=\frac{1}{2}\frac{\delta V_0}{\delta \alpha}\cdot\alpha=\frac{1}{2}\frac{i}{\lambda_B} \Sigma\cdot\alpha\ .
  	\end{aligned}
  \end{equation}
Plugging this into the equation \ref{Seff} we find that on-shell, the effective action becomes,
\begin{equation} \label{sal}
	\begin{aligned}
		S_{\text{eff}}=N_B \mathcal{V}_3 \bigg(-\frac{i}{2\lambda_B} \int 
\frac{\td^3q}{(2 \pi)^3}
\Sigma^{\nu\mu}(q) \alpha_{\mu\nu}(-q) & + \int \frac{\td^3q}{(2 \pi)^3} \log {\rm det}
(K_W(q)  -i \Sigma^{T}(q) ) \bigg)\ .
	\end{aligned}
\end{equation}
Using the formula \eqref{detQ} and the fact that $M(w)=M$ (where 
$M$ is independent of $w$), the second term on the RHS of \eqref{sal} is easily evaluated: 
\begin{equation}\label{detexp} \begin{split}
&\int \frac{\td^3q}{(2 \pi)^3} \log {\rm det}
(K_W(q)  -i \Sigma^{T}(q) ) \\
&= -\frac{|c_B|^3}{3} + 
\int_{-\pi}^{\pi} \rho(\alpha) d\alpha\int_{|c_B|}^{\infty}dy y  \left(\log(1-e^{-y  -i\alpha -\nu})+ \log(1-e^{-y+i\alpha + \nu})\right)\ ,
\end{split}
\end{equation}
(Here and the rest of the analysis in this section, we restrict
ourselves to the regime $|c_B| > |\nu|$. See Appendix \ref{review} for
a discussion). We now turn to simplifying the
first term on the RHS of \eqref{sal}. Using the fact that
$\Sigma^{33}=\Sigma^{++}=0$ (see \eqref{sigmacomp}) it follows that
that
\begin{align} \label{sigalphcont}
  &\Sigma^{\nu\mu}(q) \alpha_{\mu\nu}(-q) = \\ 
  &\quad\Sigma^{--}(q) \alpha_{--}(-q)+2( \Sigma^{-3}(q) \alpha_{3-}(-q)+\Sigma^{-+}(q) \alpha_{+-}(-q)+\Sigma^{+3}(q) \alpha_{3+}(-q) )\ .\nonumber
\end{align}
In order to further simplify \eqref{sigalphcont} we now plug in the
explicit expressions for $\Sigma$ and $\alpha$ obtained above
(i.e. \eqref{sigmacomp} and \eqref{alphaexpr} with the particular
values of $F_1 \ldots F_4$ solved for above). The dependence of the
resultant expression on $q_3$ is very simple; it is given by a
polynomial of degree one in $q_3$ times $\frac{1}{\detr Q}$.  (Of
course $q_3$ is discretized and holonomy shifted version at finite
temperature).  The linear term in this Polynomial yields a vanishing
contribution when summed over the full range of discrete values of
$q_3$ and simultaneously integrated over the holonomy \footnote{We use
  here that the eigenvalue distribution function $\rho(\alpha)$ is an
  even function of $\alpha$.}. For this reason we simply ignore the
term linear in $q_3$.  With this understanding - omitting the terms
discussed above - we have
\begin{equation}\label{sigmaalpha}
	\begin{aligned}
		&\Sigma^{\nu\mu}(q) \alpha_{\mu\nu}(-q)= -\frac{\lambda_B }{(2 \pi )^3 \text{detQ}} i m \mathcal{L}(w)\\
		\mathcal{L}(w)= &\frac{2 g(0)^3 m^2+3 \mathcal{I}(w)}{3 g(w)^2}+\frac{1}{3} \left(-2 g(0)^3 m^2-9 g(0)^2 m^2-3 \mathcal{I}(w)-6 w\right)\\ & -\frac{4}{3}  m^2 g(w)^3+m^2 g(w)^2+\frac{1}{3} g(w) \left(6 g(0)^2 m^2+4 m^2+6 w\right)
	\end{aligned}
\end{equation}
(the functions $g$ and ${\cal I}$ were defined in \eqref{defg} and
\eqref{defI} above). The dependence of \eqref{sigmaalpha} on the
discretized and holonomy shifted version of $q_3$ is entirely through
the factor of $\frac{1}{\detr Q }$. Performing the sum over the
discrete momenta in $q_3$ we find
\begin{equation}\label{procfe}
\begin{aligned}
-\frac{i}{2\lambda_B} \int  \frac{\td^3q}{(2 \pi)^3} \Sigma^{\nu\mu}(q) \alpha_{\mu\nu}(-q) &=\int \frac{\td^3q}{(2 \pi)^3}  \frac{-m}{2 (2\pi)^3 \text{detQ} }  \mathcal{L}(w)  \\
&= \frac{-m}{2 (2\pi)^3  }  \int \frac{q_s dq_s}{2\pi}
\frac{1}{\beta} \sum_{q_3} \frac{1}{-\frac{m}{(2\pi)^3}(q^2+M^2)}\mathcal{L}(w)\\
&= \frac{m}{4 \pi}  \frac{-1}{2 (2\pi)^3  } \int dw \chi(w)\mathcal{L}(w) =\frac{m}{4 \pi} \int dw \xi'(w)\mathcal{L}(w)\\
&=\frac{m }{4 \pi}\int dw \ \xi'(w) \sum_n \mathcal{L}_n(w)(-\lambda_B)^n\ .
\end{aligned}
\end{equation}
 where we have used \eqref{intdet} and \eqref{intchi}.
In the last line of \eqref{procfe} we have simply Taylor 
expanded $\mathcal{L}$ in all explicit factors of $\lambda_B$
according to the following rule. We see from \eqref{sigmaalpha} 
that $\mathcal{L}$ depends on the functions $g$ and $\mathcal{I}$. We 
use the equation \eqref{defg} to rewrite $g$ in terms of $\chi$
using
\begin{equation}
\label{gchi} 
g(w)=1+ \lambda_B\xi(w)\ ,\quad g'(w) = -\frac{\lambda_B}{2(2\pi)^3} \chi(w)\ .
\end{equation}
In a similar fashion we use \eqref{defI} to write $\mathcal{I}$ in
terms of integrals of $\xi$:
\begin{equation}
  \label{defiii}
  {\cal I}(w)= \int_0^w dz \left( 1+ \lambda_B\xi(z) \right)\ .
\end{equation}
We then Taylor expand ${\cal L}$ treating $\xi$ and all its integrals as independent of $\lambda_B$; with this understanding 
${\cal L}_n$ are defined by 
\begin{equation}
\label{mcl}
 \mathcal{L}= \sum_{n=0}^\infty \mathcal{L}_n(w)(-\lambda_B)^n.
 \end{equation}
 The various coefficient functions ${\cal L}_n$ are easily worked
 out. We find\footnote{$\mathcal{L}_0(w)=0$ is just the statement that
   this contribution is present only when interactions with gauge
   fields are turned on.}
\begin{equation} \label{lnn}
	\begin{aligned}
		 & \mathcal{L}_0(w)=0, \ \mathcal{L}_1(w)=2 m^2 \xi[0], 
		  \\&  \mathcal{L}_2(w)=-2 \mathcal{I}_\xi(w) \xi (w)-m^2 \xi (0)^2-m^2 \xi (w)^2+3 w \xi (w)^2\\
		& \mathcal{L}_3(w)=-3 \mathcal{I}_\xi(w) \xi (w)^2+4 m^2 \xi (w)^3-6 m^2 \xi (0) \xi (w)^2+2 m^2 \xi (0)^2 \xi (w)+4 w \xi (w)^3\\
		& \mathcal{L}_n(w)=\frac{1}{3} \xi (w)^{n-3} \Big(-6 n \xi (w)^2 m^2 \xi (0)-2 m^2 (n-2) \xi (0)^3+6 m^2 (n-1) \xi (0)^2 \xi (w)\\		&  \qquad\qquad +(n+1) 2 m^2 \xi(w)^3\Big) + \Big((n+1) w \xi(w)^n - n \xi(w)^{n-1} \mathcal{I}_\xi(w)\Big)\quad\text{for}\quad n\geq 4
	\end{aligned}
\end{equation}
where 
\begin{equation}
\label{ii}
 \mathcal{I}_\xi(w)=\int_0^w \xi(z) dz\ .
\end{equation}
The integral \eqref{procfe} over the last two terms in the expression
for ${\cal L}_n$, $n\geq 4$ in \eqref{lnn} can be simplified using
\begin{equation}
\begin{aligned}
dw \ \xi '(w) \left(( n+1)  \xi (w)^n w-n  \xi (w)^{n-1} \mathcal{I}_\xi(w)\right)=d( \xi (w)^{n+1} w-\xi (w)^{n} \mathcal{I}_\xi(w))
\end{aligned}
\end{equation}
It follows that the integral over those terms reduces to surface terms
which vanish in the dimensional regularization scheme\footnote{The
  fact that $\xi(\infty )|_{DR}=0$, $\mathcal{I}_\xi(0)=0$ is used
  here.}so that
\begin{equation}
  \begin{aligned}
    \int dw \ \xi '(w) \left(( n+1)  \xi (w)^n w-n  \xi (w)^{n-1} \mathcal{I}_\xi(w)\right)=0\ .
  \end{aligned}
\end{equation}
The integral over all remaining terms in ${\cal L}_n$ for all $n$ are
of the form $$\int d\xi\, f(\xi)\ ,$$ where the functions $f$ are all simple
polynomials of $\xi$.  As a consequence all remaining integrals are
easily performed and we find
  \begin{equation}
  	\begin{aligned}
  		\int dw  \ \xi '(w) \mathcal{L}_n(w)=0 \quad \text{for $n \geq 3$ }\ .
  	\end{aligned}
  \end{equation}
The only non-zero contributions are
\begin{align}
	&\int dw  \ \xi '(w) \mathcal{L}_1(w)=-2 m^2 \xi[0]^2\ ,\\
	&\int dw  \ \xi '(w) \mathcal{L}_2(w)=\frac{4}{3} m^2 \xi[0]^3\ .
\end{align}
Putting all these together we get, for the $\Sigma\cdot\alpha$ piece,
\begin{equation}\label{intp}
	\begin{aligned}
		-N_B \mathcal{V}_3 \frac{i}{2\lambda_B} \int \Sigma^{\nu\mu}(q) \alpha_{\mu\nu}(-q)=& N_B\frac{m \mathcal{V}_3}{4 \pi}(2 m^2 \xi[0]^2 \lambda_B+\frac{4}{3} m^2 \xi[0]^3 \lambda_B^2)\\
		=& N_B\frac{ \mathcal{V}_2 T^2}{6 \pi}(3  \mathcal{S}^2 |\hat{m}| |\lambda_B|+2  \mathcal{S}^3 \lambda_B^2)\ ,
	\end{aligned}
\end{equation}
where we use
$\sgn{(m)}=\sgn{(\kappa_B)}=\sgn{(\lambda_B)}$. Combining \eqref{intp}
and \eqref{detexp} we obtain
\begin{equation} \label{nnnh}
	\begin{aligned}
		&v_B(|c_B|,\rho_B) = \frac{N_B}{6 \pi}\bigg(3 |\lambda_B| |\hat{m}| \mathcal{S}^2 + 2 |\lambda_B|^2  \mathcal{S}^3  \\
&\qquad -|c_B|^3 + 3 \int_{-\pi}^{\pi} \rho_B(\alpha) d\alpha\int_{|c_B|}^{\infty}dy y \left(\log(1-e^{-y  -i\alpha -\nu})+ \log(1-e^{-y+i\alpha + \nu})\right)
\bigg)
	\end{aligned}
\end{equation}
This matches precisely with the prediction for the bosonic 
free energy from the fermionic result presented in \eqref{prdfi}. 
In other words the free energy of the bosonic theory exactly 
matches the free energy of the fermionic theory under the duality 
map, as we set out to show. 

It is not difficult to promote the expression \eqref{nnnh} to an 
offshell free energy. Consider the quantity
\begin{equation} \label{osfee}
\begin{split}
F_B[ \rho_B(\alpha),c_B] &=\frac{N_B}{6\pi} {\Bigg[}  -
\frac{ \left(\lambda_B-{\rm sgn}(\lambda_B) - {\rm sgn} (X_B) \right) }{\lambda_B}|c_B|^3 +\frac{3}{2} {\hat m}_B^{\rm cri} c_B^2 + \alpha \left({\hat m}_B^{\rm cri}\right)^3\\
&+3 \int_{-\pi}^{\pi} \rho(\alpha) d\alpha\int_{|c_B|}^{\infty}dy y\left(\ln\left(1-e^{-y-i\alpha}\right)+\ln\left(1-e^{-y+i\alpha}\right)  \right)
{\Bigg]},\\
\end{split}
\end{equation}
where $\alpha$ is an unknown pure number (see below for a 
discussion). 
In the case that  ${\rm sgn} (X_B) =-{\rm sgn}(\lambda_B)$, 
$F_B$ reduces to $v_B(\rho)$ reported in \eqref{offshellfe}.  It follows that \eqref{osfee} is the correct offshell free energy 
in the unHiggsed phase. Let us now consider the opposite case 
 ${\rm sgn} (X_B) ={\rm sgn}(\lambda_B)$. In this case the 
 expression for $F_B$ in \eqref{osfee} simplifies to 
 \begin{equation} \label{osfeeh}
 \begin{split}
 F_B[ \rho_B(\alpha),c_B] &=\frac{N_B}{6\pi} {\Bigg[}  -
 \frac{ \left(\lambda_B-2{\rm sgn}(\lambda_B) \right) }{\lambda_B}|c_B|^3 +\frac{3}{2} {\hat m}_B^{\rm cri} c_B^2+ \alpha\left({\hat m}_B^{\rm cri}\right)^3\\
 &+3 \int_{-\pi}^{\pi} \rho(\alpha) d\alpha\int_{|c_B|}^{\infty}dy y\left(\ln\left(1-e^{-y-i\alpha}\right)+\ln\left(1-e^{-y+i\alpha}\right)  \right)
 {\Bigg]},\\
 \end{split}
 \end{equation}
 
 It is not difficult to verify that the condition of stationarity 
 of variation of \eqref{osfeeh} w.r.t. $|c_B|$ yields the 
 gap equation \eqref{fge}. It is also not difficult to verify 
 that when \eqref{osfeeh} reduces to \eqref{nnnh} when evaluated
 onshell (upto the term proportional to $\alpha$: see below) i.e. on a solution to \eqref{osfeeh}. It follows 
 that \eqref{osfeeh} is an offshell free energy for the 
 critical boson theory in the Higgsed phase, and so also that 
\eqref{osfee} is an offshell free energy for the critical 
boson in either phase - Higgsed or unHiggsed. Finally, it 
is not difficult to verify that \eqref{osfee} maps to the 
regular fermionic offshell free energy reported in \eqref{offshellfe} (once we identify $|c_B|$ with $|c_F|$). 

Let us now return to a discussion of the parameter $\alpha$ in 
\eqref{osfeeh}. As this term is independent of $c_B$ it does 
not affect the variation of the action w.r.t. $c_B$ and so does 
not contribute to the gap equations. This term shifts
$\ln Z$ of the theory ($Z$ is the finite temperature partition 
function) by $- \frac{ V (m_B^{\rm cri})^3 \alpha }{T}$ where $V$ is the volume of space and $T$ is the temperature. This shift can be absorbed into a shift of the ground state energy of the theory by $V \alpha  (m_B^{\rm cri})^3$, or equivalently by a shift 
proportional to  $\alpha ( m_B^{\rm cri})^3$ of the cosmological constant counterterm of the original field theory. In other 
words the parameter $\alpha$ can only be determined once we have 
made a particular choice of the cosmological constant counterterm. In the absence of such a choice $\alpha$ is 
ambiguous. We will leave $\alpha$ above as a free parameter in 
our final result.

As we have explained in the introduction, the quantity $v_B$ reported in \eqref{nnnh} (or equivalently \eqref{prdfi})
defines the integrand of an integral over unitary matrices 
$U$. The result of this integral over $U$ is the finite 
temperature partition function $\mc{Z}$ 
\begin{equation} \label{pfh}
\mc{Z}=\text{Tr } e^{-\beta H}\ ,
\end{equation}
where $H$ is the Hamiltonian. In the Higgsed phase the Hamiltonian $H$
may be obtained by canonically quantizing the action \eqref{asb} - the
starting point of our path integral evaluation of the free energy. The
spectrum of \eqref{asb} is particularly simple in the limit
$\lambda_B=0$ with
$|m_W|=\left|\frac{\lambda_B m_B^{\text{cri}}}{2}\right|$ held
fixed. In this limit the gauge fields $A_\mu$ are very weakly coupled,
and the the partition function \eqref{pfh} may be evaluated by
enumerating the spectrum of effectively free massive $W$ (and $Z$)
bosons, subject only to the `Gauss Law' constraint that asserts that
all physical states are gauge singlets (see \cite{Aharony:2003sx} and
references therein). It is easy to see that our explicit results
\eqref{mdrfooi} and \eqref{prdfi} are consistent with this
expectation. In this limit \eqref{mdrfooi} reduces to
$|c_B|= |\hat{m}_W|$. In other words the thermal mass of the $W$
bosons agrees with their bare mass at all temperatures, as expected in
a free theory. Moreover, after dropping irrelevant constants, the
expression \eqref{prdfi} reduces, in this limit to
\begin{align} \label{prdfd}
&v_B(|c_B|,\rho_B)=\frac{N_B}{2\pi}
 \int_{-\pi}^{\pi} \rho_B(\alpha) d\alpha\int_{|\hat{m}_W|}^{\infty}dy y\left(\log\left(1+e^{-y-i\alpha -\nu}\right)+\log\left(1+e^{-y+i\alpha + \nu}\right)\right)\ ,
\end{align}
which is precisely $v_B$ of a free complex bosonic degree of freedom
(in this case the $W$ bosons) in the fundamental
representation\footnote{We thank D. Radicevic for a very useful
  discussion on this point.}.

\section{Discussion}

In this paper we have directly evaluated the thermal free 
energy of the large $N_B$ Chern-Simons gauged critical scalar 
theory in its Higgsed phase, and demonstrated that our final 
results match perfectly with the predictions of its conjectured 
fermionic dual. In particular we have demonstrated that the 
pole mass of the $W$ boson maps to the pole mass of the 
bare fermionic excitations under duality. It follows that under 
duality, the elementary fermionic excitations - which map to elementary scalar excitations in the unHiggsed phase map
to $W$ bosons in the Higgsed phase. 

At zero temperature both the bosonic theory and its fermionic dual
undergo a sharp phase transition when the bosonic/fermionic mass goes
through zero. As we have explained above, the topological pure
Chern-Simons theory that governs the long distance dynamics of the two
theories changes discontinuously from positive to negative mass, and
may be thought of as an order parameter for the phase transition.  At
finite temperature, on the other hand, there is no clear order
parameter separating the two `phases' (note in particular that the
long distance effective theory is two rather than three dimensional
and so cannot be a Chern-Simons theory). On physical grounds it seems
likely that the free energy of our theories is analytic as a function
of mass (and chemical potentials) even at finite $N$.

It is interesting that this physically expected feature of the 
free energy - namely that it is analytic as a function of mass at 
finite temperature - is borne out by the explicit large $N_B$ 
calculations presented in this paper but in a highly unusual way. 
The finite temperature free energy of the bosonic theory in its 
Higgsed `phase' is determined by a completely different computation 
than the one that determines the finite temperature free energy 
in the unHiggsed `phase'. The two calculations have non overlapping 
domains of validity, deal with different degrees of freedom and 
are dominated by distinct looking saddle points. Yet, when the 
dust settles, it turns out (in an apparently miraculous manner)
that the two results are simply analytic continuations of each other. 
At the level of formulas, therefore, there is a sense in which the duality between 
fermions and scalars is enhanced into a `triality' between fermions, 
scalars and $W$ bosons at finite temperature: there are 
three completely different looking computations, each 
of which give rise to the same final free energy after the 
appropriate analytic continuation. It would be interesting 
to understand this better - perhaps there is a more general 
uniform way of computing the bosonic free energy in both phases 
at once which makes the analyticity of the final result manifest.

From a physical point of view, the duality between bosons and fermions is particularly interesting at nonzero chemical potential and low temperatures. In this regime one expects a Fermi liquid at weak fermionic coupling but a Bose condensate at weak bosonic coupling.  By analyzing the already known fermionic results, the authors  of \cite{Geracie:2015drf} have already made this expectation 
quantitative (by dualizing the fermionic free energy to bosonic 
variables, and demonstrating that the final results at weak 
bosonic coupling enjoys certain features expected of Bose 
condensates). It would be interesting to better understand these 
results directly from the bosonic point of view using the results 
of this paper.  

The partition function of the Higgsed phase scalar theory on
$S^2 \times S^1$ is obtained by performing an integral over
holonomies; the integrand for this integral is given by the the free
energy $v_B[\rho]$ computed in this paper. From a physical point of view it
would be interesting to explore this integral in detail, particularly
at finite chemical potential. In the large volume limit the saddle
point eigenvalue distribution will take the universal tabletop form
\footnote{Given by $\rho(\alpha)=0$ for $|\alpha| > \pi |\lambda_B| $
  and $\rho(\alpha)= \frac{1}{2 \pi |\lambda_B|}$ for
  $|\alpha|< \pi |\lambda_B|$. }. However the distribution will
deviate from this universal form away from the large volume limit,
giving rise to a rich phase structure with many interesting phase
transitions (generalizing the analysis of \cite{Jain:2013py} ).

It should be possible to generalize the computations presented in this
paper to the study of the partition functions of the regular boson -
critical fermion duality (see e.g. \cite{Minwalla:2015sca}) and of
theories with with both a bosonic and a fermionic field
(\cite{Jain:2013gza, xyz}).  It would also be interesting to use the
techniques of this paper to generalize the S-matrix computations of
\cite{Jain:2014nza,Dandekar:2014era,Inbasekar:2015tsa,Yokoyama:2016sbx,Inbasekar:2017ieo,Inbasekar:2017sqp}
to evaluate the bosonic S-matrices in the Higgsed phase, and to match
the final results with the fermionic S-matrices as predicted by
duality. The techniques of this paper could also permit the
computation of the quantum effective action of the scalar theories as
a function of the gauge covariant field $\phi^a$ (in a suitable
gauge). This computation could prove useful in analysing the vacuum
stability of these theories. We hope to turn to several of these
issues in the near future.

\appendix

\acknowledgments

We would like to thank T. Sharma, T. Takimi, S. Wadia and S. Yokoyama
for collaboration during the initial stages of this project. We would
also like to thank O. Aharony, A. Gadde, D. Radicevic, and D. T. Son
for useful discussions, and O. Aharony, D. Radicevic, S. Prakash,
N. Seiberg and S. Wadia for useful comments on the manuscript.  The
work of S. C., A. D., I. H., L. J., S. M., and N. P. was supported by
the Infosys Endowment for the study of the Quantum Structure of
Spacetime. S. C. and S. J. would like to thank TIFR, Mumbai for
hospitality during the completion of the work. Finally we would all
like to acknowledge our debt to the steady support of the people of
India for research in the basic sciences.

\appendix

\section{Review of known results and a prediction for the Higgsed
	phase} \label{review}
The gap equation for the bosonic theory - which follows from varying 
\eqref{offshellfe} w.r.t. $|c_B|$ - takes the form
\begin{equation}\label{mdcb}
2 \cS(|c_B|,\nu) = {\hat m}_B^{{\rm cri} }\ ,
\end{equation}
while the gap equation for the fermionic theory is 
\begin{equation}\label{mdrf}
|c_F|=\sgn(X_F) \left( 2 \lambda_F \cC(|c_F|, \nu)  +{\hat m}_F^{{\rm reg}}\right) = |X_F|\ ,
\end{equation}
where $\cS$ and $\cC$ were defined in \eqref{ss}.

The bosonic and fermionic holonomy eigenvalue distribution functions 
are related to each other by the formula (see \cite{Jain:2013py})
\begin{equation}\label{lrtr}
|\lambda_B| \rho_B (\alpha)+ |\lambda_F| \rho_F(\pi - \alpha) = \frac{1}{2 \pi}.
\end{equation}
When \eqref{lrtr} holds (and assuming that $\rho_B(\alpha)$ and $\rho_F(\alpha)$ are even functions of their 
arguments)  it is easily verified that 
\begin{equation}\label{dualofq1} \begin{split} 
&\lambda_B \cS = \lambda_F \cC -\frac{\sgn(\lambda_F)}{2} {\rm max}(|c_F|, |\nu|)\ , %=\lambda_F \cC -\frac{\sgn(\lambda_F)}{4}\left[ ~\big| \left( |c_F| + \nu \right)  \big| + \big| \left(  |c_F| -\nu  \right) \big| ~ \right]
\\
&\lambda_F \cC=\lambda_B \cS-\frac{\sgn(\lambda_B)}{2} {\rm max}(|c_B|, |\nu|)\ . %= \lambda_B \cS-\frac{\sgn(\lambda_B)}{4}\left[ ~\big| \left( |c_B| + \nu \right)  \big| + \big| \left(  |c_B| -\nu  \right) \big| ~ \right]
\end{split}
\end{equation}
The equations \eqref{lrtr} have been derived assuming that 
the integral over $\alpha$ in the first of \eqref{dualofq1} runs 
over real $\alpha$, i.e. the unit circle in the complex plane 
$z=e^{i \alpha}.$

We pause to elaborate on the analytic structure of the functions $\cC$
and $\cS$.  The integrals over $\alpha$ in $\cC$ and $\cS$ formally
run over the range $(-\pi, \pi)$. In this paper we will, however, be
mainly interested in phases in which $\rho_F(\alpha)$ vanishes in a
neighbourhood of $\pi$ (see \cite{Jain:2013py}) for an extensive
discussion of the phases of the large $N$ partition functions of this
theory). In the rest of this paragraph we focus our attention on these
fermionic `lower gap' phases. When this is the case, it is easy to
check that the argument of the logarithmic functions that appear in
$\cC(|c_F|, \nu)$ in \eqref{ss} never pass through either zero or any
negative number for any value of $|c_B|$ or $\nu$. It follows that
$\cC(|c_F|, \nu)$ is an analytic function of its arguments for all
values of $|c_F|$ and $\nu$.

The arguments of $\cS(|c_B|, \nu)$ in \eqref{ss} are also nowhere
negative on the (unit circle) contour of integration when
$|c_B| >|\nu|$. It follows that $\cS$ is also an analytic function of
$\nu$ for $|c_B| > |\nu|$.  At $|c_B|=\nu$, on the other hand, the
arguments of one of the two logarithms in this equation goes to zero
at $\alpha=0$. For $|\nu|>|c_B|$, the contour integral passes through
the cut of the logarithm. These observations suggest that
$\cS(|c_B|, \nu)$ - viewed as a function of $\nu$ at fixed $|c_B|$ -
might well be non-analytic at $\nu=\pm |c_B|$. Equation
\eqref{dualofq1} - together with the fact that $\cC$ is analytic at
$\nu=\pm |c_B|$ - tells us that this is indeed the case. Indeed the
function $\cS$ must have precisely the singularity needed to cancel
that of the function
$\frac{\sgn(\lambda_B)}{2} {\rm max}(|c_B|, |\nu|) $ on the RHS of the
second of \eqref{dualofq1}.

\noindent The discussion of the last paragraph motivates us to define
the analytic function ${\tilde \cS}$
\begin{equation}\label{deftcs}
{\tilde \cS}= \left\{\begin{array}{cc} \cS  & {{\rm when} ~~|\nu|<|c_B|}\\ \cS - \frac{1}{2 |\lambda_B|} (|\nu|-|c_B|) & {{\rm when} ~~|\nu|>|c_B|}\end{array}\right.\ .
\end{equation}
When expressed in terms of ${\tilde \cS}$ the relations between $\cS$
and $\cC$ in \eqref{dualofq1} become the single relation
\begin{equation}\label{ccm}
\lambda_F \cC=\lambda_B {\tilde \cS}- ~\frac{\sgn(\lambda_B)}{2}|c_B|\ .
\end{equation}
Roughly speaking ${\tilde \cS}$ can be thought of as being defined by
the same integral as that for $\cS$ in the second of \eqref{ss} except
that one is instructed to perform the integral over a contour that is
deformed to avoid cutting the branch cut of the logarithmic functions.

\noindent It follows from \eqref{ccm} that under duality the quantity
$X_F = 2\lambda_F \cC + \hat{m}_F^{\text{reg}}$ defined in
\eqref{XFXC} maps to $X_B$ where
\begin{equation}\label{dbe}
X_B= 2 \lambda_B {\tilde \cS} -\lambda_B \hat{m}_B^{\text{cri}} -{\rm sgn}( \lambda_B) |c_B|\ .
\end{equation}
% \footnote{We have used 
% $$ \frac{1}{2}\left[ ~\big| \left( |c_F| + \nu \right)  \big| + \big| \left( 
% |c_F| -\nu  \right) \big| ~ \right] ={\rm max}(|c_F|, |\nu|) $$
% }
Notice that on-shell (i.e. on a solution to the bosonic gap equations)
\begin{equation}\label{xb}
X_B= -{\rm sgn}(\lambda_B)~{\rm max}(|c_B|, |\nu|)\ , ~~~
{\rm so ~~that }~~~-\lambda_B X_B \geq 0\ .
\end{equation}
In other words all solutions to the bosonic gap equations have
$\lambda_B X_B\leq 0$ i.e. $\lambda_F X_F \geq 0$. It follows
that any solution of the fermionic gap equations that violates this
inequality does not have a bosonic dual. We will now see how this
works in more detail.

\noindent Inserting \eqref{ccm} into the fermionic gap equation \eqref{mdrf} we obtain 
\begin{equation}\label{mdrff}
|c_B|=\sgn(X_B) \left( 2\lambda_B {\tilde \cS}- ~{\rm sgn} (\lambda_B)|c_B| -\lambda_B \hat{m}_B^{\text{cri}}\right),
\end{equation}
Equivalently
\begin{equation}\label{mdrfnn}
|c_B|\left( 1+{\rm sgn} (\lambda_B) \sgn(X_B) \right) =\sgn(X_B) \left( 2\lambda_B {\tilde \cS}
-\lambda_B \hat{m}_B^{\text{cri}}\right),
\end{equation}
Let us first suppose that ${\rm sgn} (\lambda_B) \sgn(X_B)=-1$. In this case \eqref{mdrfnn}
reduces to the equation
\begin{equation}\label{mbe} 
2{\tilde \cS} = \hat{m}_B^{\text{cri}}
\end{equation}
This equation matches perfectly with \eqref{mdcb} when
$|c_B|>|\nu|$\footnote{\eqref{mbe} and \eqref{mdcb} differ when
	$|\nu|>|c_B|$, because, in this regime, ${\tilde \cS}$ differs from
	$\cS$. However the difference between the two equations is quite
	minor - as we have explained above ${\tilde \cS}$ and $\cS$ are
	defined by the same integrals but over slightly different
	contours. It is possible that the derivation of \eqref{mdcb} has a
	subtlety when $|\nu|>|c_B|$ and the correct equation picks out the
	contour that changes $\cS$ to ${\tilde \cS}$. We leave an
	exploration of this to future work.}. On the other hand when
${\rm sgn} (\lambda_B) \sgn(X_B)= +1$, \eqref{mdrfnn} becomes
\begin{equation}\label{mdrfoo}
2|c_B| =\left( 2|\lambda_B| {\tilde \cS}
-|\lambda_B| \hat{m}_B^{\text{cri}}\right)\ .
\end{equation}
This is a completely new bosonic gap equation that - at least
superficially - seems different from the bosonic gap equation
\eqref{mdcb}. It has been speculated that this equations governs the
dynamics of the critical boson theory in its Higgsed phase. In the
rest of this paper we demonstrate that this is indeed the case by
directly deriving \eqref{mdrfoo} from an analysis of the bosonic
theory.

We can also use the fermionic free energy (the second of 
\eqref{offshellfe} ) together with the duality map to obtain a 
prediction for the free energy in Higgsed phase 
(we focus on the case for boson for $|c_B|>|\nu|$; when 
$|c_B|<|\nu|$ there is a potential subtlety as in the unHiggsed 
phase). 

\noindent For later use we present our results in terms of a quantity
\begin{equation}\label{convn}
m =  - \frac{\lambda_B m_B^{\text{cri}}}{2} \implies |m|= - \frac{|\lambda_B| m_B^{\text{cri}}}{2}\ ,
\end{equation}
(and correspondingly for the dimensionless hatted quantities.) Note
that in the phase under consideration $m_B^{\text{cri}}<0$. The
quantity $|m|$ would then correspond to the mass of the $W$ boson in
this phase. Using \eqref{mdrfoo} we find
\begin{equation}\label{cFtoB}
|c_B| = |\lambda_B| \cS + |\hat{m}|\ .
\end{equation}
Substituting \eqref{cFtoB} in \eqref{offshellfe}, we have (dropping
zero temperature contributions)
\begin{equation}
\begin{split}\nonumber
v_F &=\frac{N_F}{6\pi} {\Bigg[}  |c_F|^3 \frac{\left(|\lambda_F| + 1\right)}{|\lambda_F|} -\frac{3}{2|\lambda_F|} |\hat{m}_F^{\text{reg}}| c_F^2 -\\
&\qquad-3 \int_{-\pi}^{\pi} \rho_F(\alpha) d\alpha\int_{|c_F|}^{\infty}dy y\left(\log\left(1+e^{-y-i\alpha -\nu}\right)+\log\left(1+e^{-y+i\alpha + \nu}\right)  \right) {\Bigg]}, \\
&=\frac{N_B}{6\pi} {\Bigg[}  |c_B|^3 \frac{\left(2 - |\lambda_B|\right)}{|\lambda_B|} -\frac{3}{ |\lambda_B|} |\hat{m}| c_B^2 + \\
&\qquad +3 \int_{-\pi}^{\pi} \rho_B(\alpha) d\alpha\int_{|c_B|}^{\infty}dy y\left(\log\left(1+e^{-y-i\alpha -\nu}\right)+\log\left(1+e^{-y+i\alpha + \nu}\right)  \right)
{\Bigg]}\ ,\\
\end{split}
\end{equation}
\begin{align}\label{prdf}
&=\frac{N_B}{6\pi} {\Bigg[}  \frac{2 |c_B|^3 - 3|\hat{m}| c_B^2}{|\lambda_B|} + \nonumber\\
&\qquad  - |c_B|^3 + 3 \int_{-\pi}^{\pi} \rho_B(\alpha) d\alpha\int_{|c_B|}^{\infty}dy y\left(\log\left(1+e^{-y-i\alpha -\nu}\right)+\log\left(1+e^{-y+i\alpha + \nu}\right)  \right)
{\Bigg]}\nonumber\\
&=\frac{N_B}{6\pi} {\Bigg[}-\frac{|\hat{m}|^3}{|\lambda_B| }+3|\lambda_B|   |\hat{m}| \cS^2+2|\lambda_B| ^2  \cS^3 \nonumber\\
&\qquad + 3 \int_{-\pi}^{\pi} \rho_B(\alpha) d\alpha\int_{|c_B|}^{\infty}dy y\left(\log\left(1+e^{-y-i\alpha -\nu}\right)+\log\left(1+e^{-y+i\alpha + \nu}\right)\right)  -|c_B|^3  {\Bigg]}\ ,
\end{align}
where we have used the following duality maps in the first step:
\begin{equation}
\begin{split}
\hat{m}_F^{\text{reg}} = 2 \hat{m}\ ,\quad \frac{N_F}{|\lambda_F|} =
\frac{N_B}{|\lambda_B|}, \quad |\lambda_F| = 1-|\lambda_B|\ ,\quad
|\lambda_F| \rho_F(\alpha) = \frac{1}{2\pi} - |\lambda_B| \rho(\pi -
\alpha)\ .
\end{split}
\end{equation}
In the second and third steps, we have rearranged terms in the
expression in order to put it in a form which will match term by term
with the free energy obtained by direct calculation in the Higgsed
phase. The first term in the bracket is a zero temperature
contribution and can be ignored in this context.

% BIBLIOGRAPHY
% use BIBTEX if you want
%\bibliographystyle{JHEP}
%\bibliography{yourBIBfiles}

% The bibliography will probably be heavily edited during typesetting.
% We'll parse it and, using the arxiv number or the journal data, will
% query inspire, trying to verify the data (this will probalby spot
% eventual typos) and retrive the document DOI and eventual errata.
% We however suggest to always provide author, title and journal data:
% in short all the informations that clearly identify a document.

\bigskip
%\bibliography{cp.bib}

\providecommand{\href}[2]{#2}\begingroup\raggedright\endgroup

\end{document}